\documentclass[aps,prb,twocolumn]{revtex4} %,preprint

\usepackage{graphicx}% Include figure files
\usepackage{dcolumn}% Align table columns on decimal point
\usepackage{bm}% bold math

\newcommand{\comment}[1]{}
\begin{document}
\renewcommand{\theequation}{\arabic{section}.\arabic{equation}}
%\preprint{}   % Preprint number in upper right corner

\title{Optimising Principle
for Non-Equilibrium Phase Transitions
and Pattern Formation
with Results for Heat Convection}

%\author{}
%\thanks{}

\author{Phil Attard}
%\affiliation{\protect\texttt{phil.attard1@gmail.com}}

\date{25 July, 2012. phil.attard1@gmail.com}

\begin{abstract}
Spontaneous transitions between non-equilibrium patterns
are characterised by hydrodynamic calculations
of ideal straight roll steady state heat convection.
The calculations are tested quantitatively
against existing experimental data.
It is shown that at a given Rayleigh number
the final wave number depends upon whether the initial state
was the conducting state,
or, in the case of the cross roll transition,
the wave number of the initial straight roll state.
The final wave number does not correspond
to the maximum or to the minimum heat flux (entropy production or dissipation),
nor to the maximum sub-system entropy.
In all cases the entropy of the total system increases
monotonically during the spontaneous transition.
It is concluded that there does not exist
any single time thermodynamic property or variational principle
for non-equilibrium systems.
It is further concluded that the second entropy is \emph{the}
two time variational principle
that determines the optimum non-equilibrium state or pattern.
\end{abstract}

%\keywords{}

\maketitle

%%%%%%%%%%%%%%%%%%%%%%%%%%%%%%%%%%%%%%%%%%%%%%%%%%%%%%%%%%%%%%%%%%%%%%%%%%%%%%%
%                                                                             %
                \section{Introduction}
%                                                                             %
%%%%%%%%%%%%%%%%%%%%%%%%%%%%%%%%%%%%%%%%%%%%%%%%%%%%%%%%%%%%%%%%%%%%%%%%%%%%%%%

In the case of equilibrium systems,
there is a well established variational principle
for determining the optimum state,
namely that the total entropy is a maximum,
or, equivalently, that the free energy is a minimum.
In the case of non-equilibrium systems,
there is no such generally agreed principle,
despite much effort and many claims.
The two most common ideas invoke the rate of entropy production,
with one school of thought claiming that it is a maximum,
and another school of thought claiming that it is a minimum.
The postulated ideas are mutually contradictory
and there is a dearth of experimental or computational evidence
to support either view.

In what remains of this introduction,
and to foreshadow the numerical results for heat convection
that are the main work of this paper,
it is perhaps of historical and psychological interest
to explore why the rate of entropy production
continues to be championed as a non-equilibrium variational principle
despite the acknowledged lack of predictive success of the approach.
In the first place there is the implicit assumption
that their \emph{ought} to be an optimum non-equilibrium state.
To question this assumption is not so bizarre as it might seem at first sight,
as  the numerical results obtained in this paper will show.
The reason for uncritically assuming
the existence of a non-equilibrium variation principle
probably lies in the observation that in a given non-equilibrium system
there are often quite distinct states or patterns
that spontaneously occur,
often reproducibly influenced by some control parameter.
For example, in heat flow,
there is the critical Rayleigh number at which the conduction--convection
transition occurs,
and for convection,
there appear distinct steady state patterns
(ideal straight rolls, hexagonal cells)
of particular wave numbers, depending upon the boundary conditions,
and other sometimes unidentified influences.
The patterns that form in non-equilibrium systems
can be quite spectacular,
and it is natural to assume that the one selected by the system
in a given case is the result of some general thermodynamic principle.

This idea that there must exist a variational principle
for non-equilibrium systems
is also motivated by the success of the Second Law of Thermodynamics
for equilibrium systems.
The  principle
that the optimum equilibrium state is the state of maximum entropy
has been demonstrably successful in accounting for the selection
of competing equilibrium states.
One can see how this might motivate many to think
that there ought to exist an analogous variational principle
for non-equilibrium systems.
From this many workers simply extrapolate
the Second Law of (Equilibrium) Thermodynamics to assert that,
since in the time-independent case entropy is a maximum,
in the time-dependent case it is the rate of production
of entropy that must be a maximum.
Whereas the Second Law of (Equilibrium) Thermodynamics
was postulated by Clausius \emph{after}
much experimental observation and measurement,
the principle that the rate of entropy production
be maximised does not come from any observation, calculation, or measurement.
What is more, Boltzmann's discovery of the molecular basis
of entropy in terms of the number of configurations in a state
provided a rational explanation, a physical interpretation,
and mathematical framework  for the Second Law.
This again contrasts with the non-equilibrium case
where there is a disconnect between the molecular picture of entropy
and the assertion that the rate of entropy production is a maximum.

The historical origin of the opposite postulate,
that the rate of entropy production is a minimum,
appears to predate thermodynamics itself.
As discussed by Jaynes
in his review of the
Principle of Minimum Entropy Production,\cite{Jaynes80}
in 1848 Kirchhoff\cite{Kirchhoff48}
developed certain theorems for electrical circuits,
and showed that the electrical currents that flowed
could be derived from a Principle of the Least Dissipation of Energy.
Of course there is no question of the mathematical veracity
of Kirchoff's theorem for electrical circuits,
but, as is discussed next,
there is considerable reason to doubt the thermodynamic interpretation
that they are equivalent
to a Principle of Least Dissipation of Entropy
and that such a principle has general application for non-equilibrium systems.

In \S\ref{Sec:Max-Min-Diss} that follows this introduction,
a detailed historical review and mathematical  critique
of these two conventional ideas for non-equilibrium systems is given.
As well,  comparison is briefly made with the author's
own variational principle for non-equilibrium systems,
namely that the second entropy is maximised.
In \S\ref{Sec:convheatflow},
the hydrodynamic equations for convection are given,
and in  \S\ref{Sec:s1}
expressions for the total first entropy of convection are derived.
The computer algorithm
is given in  \S\ref{Sec:algo-strt}
for the one dimensional case of ideal straight rolls,
and in  \S\ref{Sec:algo-cross}
the two dimensional case of the cross roll transition.
In \S\ref{Sec:Results},
the convection calculations are compared with experimental measurements
for spontaneous transitions,
and various properties of heat convection are tested.
In \S\ref{Sec:Concl}
general conclusions are drawn from the numerical results
for the optimising principle
that gives the selected state of non-equilibrium systems.

%%%%%%%%%%%%%%%%%%%%%%%%%%%%%%%%%%%%%%%%%%%%%%%%%%%%%%%%%%%%%%%%%%%%%%%%%
%
\section{Critique of Non-Equilibrium Variational Principles}
\label{Sec:Max-Min-Diss}
%
%%%%%%%%%%%%%%%%%%%%%%%%%%%%%%%%%%%%%%%%%%%%%%%%%%%%%%%%%%%%%%%%%%%%%

In this section three non-equilibrium variational principles,
---Minimum Dissipation, Maximum Dissipation, and Second Entropy---
are presented in their simplest forms
in order to compare their strengths and weaknesses,
and the evidence for and against each one.

%%%%%%%%%%%%%%%%%%%%%%%%%%%%%%%%%%%%%%%%%%%%%%%%%%%%%%%%%%%%%%%%%%%%%%%%%
\subsection{Principle of Minimum Dissipation}

%%%%%%%%%%%%%%%%%%%%%%%%%%%%%%%%%%%%%%%%%%%%%%%
\subsubsection{Kirchhoff's Version}

Onsager,\cite{Onsager31a}
in his celebrated paper on the reciprocal relations,
gave two different Principles of Minimum Dissipation.
The first  is directly related to Kirchhoff's
Principle of Least Energy Dissipation and is treated here.
The second is related to the vanishing of fluxes
conjugate to quiescent variables
and is discussed in the following sub-section.

In 1848 Kirchhoff generalised Ohm's law to three dimensions
and noted that
the current in any region was determined
by minimising the energy dissipation
(Joule heating) with respect to the voltage
for specified values on the surface.\cite{Kirchhoff48}
This principle of least energy dissipation
was derived by Kirchhoff from the fact that
in the steady state electrical charge could not accumulate at any point.

For the present purposes,
and following Jaynes,\cite{Jaynes80}
it is sufficient to consider two resistors,
$R_1$ and $R_2$  connected in parallel
with total current $J = J_1 + J_2$ flowing through them.
(The total current is equivalent to a potential drop of
$\phi = J [R_1^{-1} + R_2^{-1}]^{-1}$.)
The energy dissipation or Joule heat is
\begin{equation}
\dot E
= J_1^2 R_1 + J_2^2 R_2
= J_1^2 R_1 + [J-J_1]^2 R_2 .
\end{equation}
The derivative of this at constant total current
(equivalently, fixed potential drop) is
\begin{equation}
\frac{\partial \dot E }{\partial J_1}
= 2 J_1 R_1 - 2 [J-J_1] R_2 ,
\end{equation}
which vanishes when
\begin{equation}
\overline J_1 = \frac{R_2}{R_1+R_2}J  .
\end{equation}
This current minimises the energy dissipation
and is of course just the current that one would obtain
directly from Ohm's law.

The Principle of Least Entropy Dissipation
appears to follow from this by considering the case
that the resistors are in thermal contact
with a reservoir of temperature $T$,
so that the rate of production of entropy in the reservoir
is just $\dot S_\mathrm{r} = \dot E/T$.
Hence minimising the energy dissipation
is the same as minimising the entropy dissipation.

There are two major problems with the extrapolation
of Kirchhoff's principle of electrical circuits
to a general principle for non-equilibrium systems.
First, the principle of least dissipation of energy
has nothing to do with entropy.
To see this simply consider the case
that the resistors are in contact with two separate reservoirs
of temperature $T_1$ and $T_2$ respectively.
The currents do not change,
but the entropy production is no longer a minimum.
(The minimum entropy dissipation would occur
when $\overline J_1 = R_2J/T_2[{(R_1/T_1)+(R_2/T_2)}] $,
which obviously violates Ohm's law.)

The first problem also signifies a common misunderstanding
of non-equilibrium variational principles,
namely it is often not precisely clear what one is trying to achieve
by such a principle.
In the present problem of current flow,
Ohm's law already completely specifies the current flow for the stated problem.
There is no point in formulating a variational function
for current flow unless it has application beyond Ohm's law.
More generally,
any non-equilibrium variational principle
has to be explicitly formulated as consistent with,
but providing something beyond, the linear transport laws.
This point will be further discussed below.

The second problem with the thermodynamic interpretation
is that only part of the entropy dissipation has been considered.
The battery or source of electromotive force
that drives the current also dissipates entropy.
If the state of non-equilibrium systems were truly determined
by the entropy dissipation,
then surely it is the total entropy dissipation that is relevant.

On this point it is worth mentioning the work of
\v{Z}upanovi\'{c} et al.,\cite{Zupanovic04}
also based on Kirchhoff's laws,
but in this case giving  a Principle of \emph{Maximum} Entropy Dissipation.
The reason that they obtained a maximum rather than
the more traditional minimum
is that they took into account
the energy supplied by the  electromotive forces
and invoked energy conservation as a constraint
on the variation of the Joule heat.
But their thermodynamic conclusion
is vitiated for the same reason as just discussed,
namely that the result comes directly from Ohm's law
(and energy conservation)
and has nothing to do with entropy.
If the resistors in the circuit are at different temperatures,
and if the entropy dissipated by the batteries and sources of
electromotive force is taken into account,
then the entropy dissipation would no longer be maximal.

Onsager,
in the first of his two celebrated papers on the reciprocal relations,%
\cite{Onsager31a,Onsager31b} gives a non-equilibrium variational
principle, Eq.~(6.6) of Ref.~\onlinecite{Onsager31a},
which case he says has been named the
`Principle of the Least Dissipation of Energy'.
Onsager says that if the boundary fluxes are specified,
and if the sub-system is in a steady state,
then Rayleigh's dissipation function, a quadratic of the fluxes, is a minimum.
Rayleigh's dissipation function has the same appearance as the Joule heat in
Kirchhoff's Principle of Least Dissipation of Energy.
Four criticisms can be made of Onsager's proposal,
all of which can be made of any thermodynamic interpretation
of Kirchhoff's Principle:
First, that Rayleigh's dissipation function does not
give the actual entropy dissipation for the arbitrary fluxes
that are explored in the variational procedure.
Second, that the rate of total entropy production
is not accounted for.
Third, since one can multiply Rayleigh's dissipation function
by an arbitrary constant,
including a negative one, and still obtain the same optimum value,
the variational function cannot be a physical thermodynamic potential.
Fourth, and connected with the third,
the variational procedure yields nothing beyond the
built-in linear transport laws,
not even the reciprocal relations.

%%%%%%%%%%%%%%%%%%%%%%%%%%%%%%%%%%%%%%%%%%%%%%%%%%
\subsubsection{Onsager's Version}

Onsager gives a second form
of the Principle of Minimum Entropy Dissipation
(Principle of Least Dissipation, for short)
in \S~6 of Ref.~\onlinecite{Onsager31a},
leading up to and following his Eq.~(6.5).
This version of the Principle
is said to apply to certain non-equilibrium systems
when some of the fluxes vanish.
Identical Principles of Minimum Entropy Dissipation
for vanishing fluxes
have been invoked by Prigogine,\cite{Prigogine67}
and by de Groot and Mazur.\cite{deGroot69}

Onsager claims that this Principle gives rise to the reciprocal relations
(see p.~424 of Ref.~\onlinecite{Onsager31a}).
He dwells upon the historical antecedents of the Principle
(he attributes it to Helmholtz and to Kelvin rather than to Kirchhoff)
in his acceptance speech for the Nobel prize.\cite{Onsager-Nobel}
The claim that it is the basis of the reciprocal relations
is also made by Mazur in his summary of Onsager's achievements.\cite{Mazur96}
If true, such a result would provide a distinct
derivation of the reciprocal relations
that was independent of the microscopic reversibility derivation
that Onsager actually used,
and it would arguably justify regarding
this particular Principle of Minimum Entropy Dissipation
as a fundamental thermodynamic principle.
This claim will now be tested.

For simplicity consider an isolated system,
with ${\bf x} \equiv \{x_1, x_2, \ldots , x_n \}$
being the non-conserved material quantities of interest.
The associated fluxes are ${\bf J} \equiv \{J_1, J_2, \ldots , J_n \}$
with $J_i \equiv \dot x_i^0/V$,
the superscript zero signifying an isolated system.
The conjugate thermodynamic forces,
${\bf X} \equiv \{X_1, X_2, \ldots , X_n \}$,
are the derivatives of the entropy,
$X_i \equiv \partial S({\bf x})/\partial x_i$.
For example,
the relevant extensive material variable
might be the first energy moment,
in which case the thermodynamic force
is the reciprocal of the first temperature,
which is essentially the temperature gradient,
and the flux is the energy crossing a plane per unit area per unit time.

The specified structure ${\bf x}$,
or equivalently force ${\bf X}({\bf x})$,
is considered to have arisen as a fluctuation
from the equilibrium state  of the isolated system,
and the flux ${\bf J}$ represents
the regression of the fluctuation back to the equilibrium state.

It will turn out that the analysis is equivalent to examining the
coupling between forces and fluxes in the case that some of the
variables are relevant variables, and others are quiescent variables.
Relevant or active variables have a specified value of the
thermodynamic force
(equivalently the value of the material property),
that results either from a fluctuation of an isolated system
or else from exchange with reservoirs.
Quiescent or slave variables have
either zero force
(in the instant of the initial fluctuation
of the active variable for an isolated system),
or zero flux,
(for a steady state sub-system with
the quiescent variables being non-exchangeable with the reservoir).

For example, in the case of thermodiffusion,
one has two material properties,
namely the moments of energy and number.
If one wants to concentrate on heat flow,
then one can still use the two component formalism,
but treat the number variables as quiescent
and set their fluxes to zero
(and one can calculate the non-zero value of the
thermodynamic force on the number).
Alternatively, and ultimately equivalently,
one can consider a single component of energy alone,
and never consider explicitly the number fluxes or forces.
In the full case when both are active,
the energy and mass fluctuations are non-zero,
and their fluxes are both non-zero.

There are two questions:
First,
do the reciprocal relations imply
the Principle of Least Dissipation
(with respect to the quiescent fluxes)?
Second,
does the Principle of Least Dissipation
imply the reciprocal relations?

For simplicity consider everything to be uniform in space.
By definition,
the rate of entropy production per unit volume of the system is
\begin{equation} \label{Eq:dotS-JX}
\dot S/V \equiv {\bf J} \cdot {\bf X} .
\end{equation}
This exact formula holds whether or not the fluxes are the optimal ones.
This will be called the dissipation.

The phenomenological or linear transport laws
say that the fluxes are proportional to the thermodynamic forces,
\begin{equation}
{\bf J} = \underline{ \underline A} {\bf X} .
\end{equation}
In this form of the linear transport laws,
the $n$ forces are regarded
as the independent variables.

Here Onsager's notation for the fluxes is used
rather than the present author's.
The present author always distinguishes
between arbitrary fluxes ${\bf J}$
and optimal fluxes $\overline {\bf J}$;
only the latter obey the linear transport laws.
Onsager (and most other authors)
do not distinguish the two explicitly.
Because one is carrying out variational procedures
for arbitrary fluxes,
it can be difficult to figure out precisely
which equations assume the linear transport laws
and which do not.

With the linear transport laws
in this form with the forces as independent variables,
the dissipation becomes a quadratic form,
\begin{equation}
\dot S/V =  {\bf X} \cdot \underline{ \underline A} {\bf X} .
\end{equation}
Since the Second Law of Thermodynamics
mandates that the dissipation be positive,
this implies that
the transport matrix must be positive definite.

The Onsager reciprocal relations says that
the transport matrix is symmetric,
\begin{equation}
\underline{ \underline A} = \underline{ \underline A}^\mathrm{T}.
\end{equation}
To answer the  first question this will be assumed true.
To answer the  second question it will be checked
whether or not the Principle of Least Dissipation
implies this result.

Now the first $m$ variables will be considered active,
and the second $n-m$ will be considered quiescent.
Use a sub-script 1 for the first $m$ variables
and 2 for the second  $n-m$  variables,
so that ${\bf J}_1 \equiv \{J_1, J_2, \ldots , J_m \}$
and  ${\bf J}_2 \equiv \{J_{m+1}, J_{m+2}, \ldots , J_n \}$,
and similarly for ${\bf X}_1$ and ${\bf X}_2$.
The transport matrix $\underline{ \underline A}$
similarly breaks into four blocks,
and the reciprocal relations imply
\begin{equation} \label{Eq:rec-rel-block}
\underline{ \underline A}_{11}  = \underline{ \underline A}_{11}^\mathrm{T}
, \;
\underline{ \underline A}_{22}  =
\underline{ \underline A}_{22}^\mathrm{T}
, \mbox{ and }
\underline{ \underline A}_{12}  =
\underline{ \underline A}_{21}^\mathrm{T} .
\end{equation}
In block form, the linear transport laws are explicitly
\begin{eqnarray} \label{Eq:Trans-Block}
{\bf J}_1
& = &
\underline{ \underline A}_{11} {\bf X}_1
+
\underline{ \underline A}_{12} {\bf X}_2
\nonumber \\
{\bf J}_2
& = &
\underline{ \underline A}_{21} {\bf X}_1
+
\underline{ \underline A}_{22} {\bf X}_2 .
\end{eqnarray}

Choosing the first $m$ forces and the second $n-m$ fluxes
as the independent variables,
the linear transport laws may be rewritten
\begin{eqnarray} \label{Eq:Trans-J1X2}
{\bf J}_1
& = &
\left[ \underline{ \underline A}_{11}
-
\underline{ \underline A}_{12}
\underline{ \underline A}_{22}^{-1} \underline{ \underline A}_{21}
\right] {\bf X}_1
+
\underline{ \underline A}_{12} \underline{ \underline A}_{22}^{-1} {\bf J}_2
\nonumber \\
{\bf X}_2
& = &
-\underline{ \underline A}_{22}^{-1} \underline{ \underline A}_{21} {\bf X}_1
+ \underline{ \underline A}_{22}^{-1} {\bf J}_2  .
\end{eqnarray}
The rate of entropy production
is a quadratic form in these independent variables
\begin{eqnarray} \label{Eq:Sdot-gen}
\dot S/V & = &
{\bf J}_1 \cdot {\bf X}_1
+
{\bf J}_2 \cdot {\bf X}_2
\nonumber \\ & = &
{\bf X}_1 \cdot \left[ \underline{ \underline A}_{11}
-
\underline{ \underline A}_{12}
\underline{ \underline A}_{22}^{-1} \underline{ \underline A}_{21}
\right]  {\bf X}_1
+
{\bf X}_1 \cdot
\underline{ \underline A}_{12} \underline{ \underline A}_{22}^{-1}
{\bf J}_2
\nonumber \\ & & \mbox{ }
- {\bf J}_2 \cdot
\underline{ \underline A}_{22}^{-1} \underline{ \underline A}_{21}
{\bf X}_1
+ {\bf J}_2 \cdot
\underline{ \underline A}_{22}^{-1}
{\bf J}_2
\nonumber \\ & = &
{\bf X}_1 \cdot \underline{ \underline A}^{(m)} {\bf X}_1
+ {\bf J}_2 \cdot
\underline{ \underline A}_{22}^{-1}
{\bf J}_2
\nonumber \\ & & \mbox{ }
+ {\bf J}_2 \cdot
\left[
( \underline{ \underline A}_{22}^{-1} )^\mathrm{T}
\underline{ \underline A}_{12}^\mathrm{T}
-
\underline{ \underline A}_{22}^{-1} \underline{ \underline A}_{21}
\right] {\bf X}_1 .
\end{eqnarray}
Here $\underline{ \underline A}^{(m)}
\equiv  \underline{ \underline A}_{11}
-
\underline{ \underline A}_{12}
\underline{ \underline A}_{22}^{-1} \underline{ \underline A}_{21} $
is the $m \times m$ transport matrix
that would hold if the active components only were considered.
In this case the linear transport laws are
${\bf J}_1 = \underline{ \underline A}^{(m)} {\bf X}_1$,
as will now be shown.

Now \emph{if} the reciprocal relations hold,
then Eq.~(\ref{Eq:rec-rel-block}) implies
\begin{equation} \label{Eq:rec-rel-block-cons}
( \underline{ \underline A}_{22}^{-1} )^\mathrm{T}
\underline{ \underline A}_{12}^\mathrm{T}
=
\underline{ \underline A}_{22}^{-1} \underline{ \underline A}_{21} .
\end{equation}
Hence the cross term vanishes
and the dissipation becomes
\begin{equation} \label{Eq:Sdot-J1J2}
\dot S/V =
{\bf X}_1 \cdot
%\left[ \underline{ \underline A}_{11} - \underline{ \underline A}_{12}
%\underline{ \underline A}_{22}^{-1} \underline{ \underline A}_{21} \right]
\underline{ \underline A}^{(m)}
{\bf X}_1
+ {\bf J}_2 \cdot
\underline{ \underline A}_{22}^{-1}
{\bf J}_2 .
\end{equation}
Because the full transport matrix is positive definite,
each of the two transport matrices that appear here
must also be positive definite
due to the combination of sub-blocks
from which they are formed.
Hence one can see that the reciprocal relations imply
that the dissipation is minimised by the vanishing of the quiescent fluxes
\begin{equation}
\left.
\frac{\partial \dot S/V }{\partial {\bf J}_2 }
\right|_{{\bf J}_2  = {\bf 0}} = {\bf 0}.
\end{equation}
In this case, because ${\bf J}_2  = {\bf 0}$,
the linear transport laws, Eq.~(\ref{Eq:Trans-J1X2}),
become
${\bf J}_1 = \underline{ \underline A}^{(m)} {\bf X}_1$,
which are those that would apply if only the active components
were considered.

This result proves that if the reciprocal relations hold,
then the entropy dissipation is minimised
by the vanishing of the quiescent fluxes.
The significance or otherwise of this result will be discussed shortly.

Now to the second question:
does  minimising the dissipation with respect to the
quiescent fluxes imply the reciprocal relations?
Differentiating with respect to the independent fluxes
the form of the dissipation
that does not assume the reciprocal relations,
Eq.~(\ref{Eq:Sdot-gen}),
one obtains
\begin{equation}
\frac{\partial \dot S/V }{\partial {\bf J}_2 }
=
\left[
( \underline{ \underline A}_{22}^{-1} )^\mathrm{T}
\underline{ \underline A}_{12}^\mathrm{T}
-
\underline{ \underline A}_{22}^{-1} \underline{ \underline A}_{21}
\right]
{\bf X}_1
+
2 \underline{ \underline A}_{22}^{-1} {\bf J}_2 .
\end{equation}
Demanding this vanish at ${\bf J}_2 = {\bf 0}$
yields Eq.~(\ref{Eq:rec-rel-block-cons}),
\[
( \underline{ \underline A}_{22}^{-1} )^\mathrm{T}
\underline{ \underline A}_{12}^\mathrm{T}
=
\underline{ \underline A}_{22}^{-1} \underline{ \underline A}_{21} .
\]
But this equation does not in general imply
the block form of the reciprocal relations,  Eq.~(\ref{Eq:rec-rel-block}),
since the matrices cannot be separately equated.

For the case $n=2$ and $m=1$ these matrices are scalars.
In this case, and only in this case,
the scalar $A_{22}$ cancels both sides
and one is left with the scalar equality $A_{12} = A_{21}$,
which is the reciprocal relation for two coupled flows.
It is to be noted that both Onsager
(see the discussion following Eq.~(6.5) of Ref.~\onlinecite{Onsager31a}),
and Mazur
(see the equations and discussion leading to
Eq.~(7) of Ref.~\onlinecite{Mazur96}),
in claiming that the Principle of Least Dissipation implies
the reciprocal relations,
both offer only an $n=2$, $m=1$ example.
One cannot deduce the properties
of the matrix $\underline{ \underline A}$ for general $n$
from the pairwise results obtained for $n=2$
without making the additional assumption
that the individual elements do not depend upon
the other components of the system,
which in any case appears to be false.
For example,
it is not true that the Soret coefficient for thermodiffusion
does not depend on the specific solvent
or on the concentrations of the other solutes.
Whereas the reciprocal relations imply
$A_{ij}^{(n)} = A_{ji}^{(n)}$,
the extrapolation of the pairwise result
would require $A_{ij}^{(n)} = A_{ji}^{(n')}$,
which is not true in general.

In contrast, the reciprocal relations hold for
an arbitrary number of active and quiescent variables;
Eq.~(\ref{Eq:rec-rel-block}) contains more information
than  Eq.~(\ref{Eq:rec-rel-block-cons}).
One concludes that the reciprocal relations
are sufficient but not necessary to ensure
that the dissipation is minimised when the quiescent fluxes vanish.
The present author cannot agree with the claim by Onsager\cite{Onsager31a}
and by Mazur\cite{Mazur96}
that the Principle of Least Dissipation for quiescent fluxes
implies the reciprocal relations.

Finally then,
what actually is the use of the Principle of  Minimum Dissipation?
Is it really so surprising or significant that the entropy dissipation
is reduced by setting some of the fluxes to zero,
Eq.~(\ref{Eq:Sdot-J1J2})?
Is it even true that in the actual physical problem%
---the regression of a fluctuation of an isolated system---%
the flux of quiescent variables vanishes?
In fact, at the first instant $\tau=0$ of a fluctuation
in the active variables, the force and flux of the quiescent
variables is zero.
At some intermediate time after that,
the quiescent force is non-zero,
and so the quiescent flux also must have been non-zero leading up to that state,
and it must be non-zero going forward in time
as both active and quiescent forces relax back to zero.
(This is different to the steady state situation
for a sub-system that can exchange with reservoirs,
in which case the active and quiescent forces are constant in time,
and the quiescent flux is zero.)
%Hence for the regression of a fluctuation of an isolated system
%the dissipation is \emph{almost never} a minimum
%with respect to the quiescent flux.
One can conclude that whilst the reciprocal relations
imply that the dissipation is minimised when the quiescent fluxes vanish,
in many cases in the real world the quiescent fluxes do not vanish
and the dissipation is not a minimum with respect to them.

More generally,
the Principle of  Minimum Dissipation does not give
either the values of the active fluxes
or the quiescent forces,
since these come from the linear transport laws
for the given active forces and the zero quiescent fluxes,
Eq.~(\ref{Eq:Trans-Block}).
It is difficult to identify anything of import
that  is achieved by minimising the dissipation
with respect to the quiescent fluxes,
or to see how one might erect a full thermodynamic theory
for non-equilibrium systems on this Principle.

%%%%%%%%%%%%%%%%%%%%%%%%%%%%%%%%%%%%%%%%%%%%%%%%%%%%%%%%%%%%%
\subsection{Principle of Maximum Dissipation}

Although most who assert the Principle of Maximum Dissipation
do so based on nothing more than
an analogy with the Second Law of Equilibrium Thermodynamics,
Onsager actually attempted a detailed justification.
Onsager believed that a particular function that he formulated,
the entropy dissipation less Rayleigh's dissipation function,
was a thermodynamic potential for non-equilibrium systems.
In fact, he asserted that
`the reciprocal relations\ldots can be expressed in terms of a potential,
and permit the formulation of a variational principle',
(just prior to Eq.~(5.1) of Ref.\onlinecite{Onsager31a}).

Onsager,
in Eq.~(5.3) of Ref.~\onlinecite{Onsager31a},
defined what will be called here the Rayleigh dissipation function,
\begin{equation}
\phi({\bf J}) \equiv
\frac{\alpha}{2} {\bf J} \cdot \underline{\underline A}^{-1} {\bf J} .
\end{equation}
This is written in the present notation,
with the constant $\alpha$ being introduced here
in order to make a point about the arbitrariness
of Onsager's functional; Onsager implicitly chooses $\alpha = 1$.
Onsager says that this dissipation function
was originally introduced by Rayleigh
as a potential for frictional forces.\cite{Rayeigh73}

As mentioned above,
there is a certain ambiguity in Onsager's work
regarding whether he regards the fluxes as arbitrary,
as here and in his Eq.~(5.3),\cite{Onsager31a}
or whether he restricts them to satisfying
the linear transport laws,
which he does in the second expression that he gives
for the Rayleigh dissipation function, Eq.~(5.5).\cite{Onsager31a}

This ambiguity underscores a significant problem
with Onsager's work,
namely  that the Rayleigh dissipation function
is only related to the entropy dissipation
(it is equal to $\alpha/2$ times the entropy dissipation)
when the fluxes are equal to the values given by the linear transport laws.
For arbitrary fluxes, this function is  not directly related
to the entropy dissipation.
Since Onsager is developing a variational principle for arbitrary fluxes,
which he believes has the physical meaning
of the non-equilibrium thermodynamic potential,
it is crucial to appreciate that Rayleigh's dissipation function
does not have the physical
interpretation given it by Onsager.

Onsager, in Eq.~(5.9) of Ref.~\onlinecite{Onsager31a},
shows that the rate of total entropy production per unit volume,
(here assuming spatial homogeneity), is
\[
{\dot S}/{V} = {\bf J} \cdot  {\bf X},
\]
which was given above as Eq.~(\ref{Eq:dotS-JX}).
Although Onsager derived this for the general case
of a sub-system and a reservoir,
it also holds for the total isolated system considered here.

Onsager gives his variational functional as
\begin{eqnarray} \label{Eq:Onsager-vary}
{\cal O}({\bf J}|{\bf X})
& \equiv &
\frac{\alpha \dot S}{V} - \phi({\bf J})
\nonumber \\ & = &
\alpha {\bf J} \cdot  {\bf X}
-\frac{\alpha}{2} {\bf J} \cdot \underline{\underline A}^{-1} {\bf J} ,
\end{eqnarray}
where again the $\alpha$ has been inserted here,
since Onsager actually took $\alpha = 1$.
It was of course necessary for Onsager to subtract
the quadratic dissipation function $\phi({\bf J})$
from the  actual entropy dissipation
because  $\dot S$
is linear in the flux and does not have an extremum.
(A common feature of all the variational principles
is that they must contain a term quadratic in the fluxes
and a term linear in the fluxes,
with the latter being proportional to the entropy dissipation.)
Onsager says that his function ${\cal O}({\bf J}|{\bf X})$
is to be maximised
over the fluxes for specified forces,
(since the dissipation must be positive,
$\underline{\underline A}$ must be positive definite,
and the extremum has to be a maximum).

Onsager offers no particular physical justification for this choice of
functional other than that it is optimised by the linear transport laws,
\begin{equation}
\overline {\bf J} = \underline{\underline A}  {\bf X}.
\end{equation}
The maximum value of the variational functional is
\begin{equation}
{\cal O}(\overline{\bf J}|{\bf X})
= \phi(\overline{\bf J})
= \frac{\alpha}{2} \overline{\bf J} \cdot  {\bf X} .
\end{equation}
The right hand side is just  $\alpha/2$
times the rate of entropy dissipation in the optimal state.

The linear transport laws emerge irrespective of the value of $\alpha$.
Onsager's analysis is equivalent to choosing
a value of $\alpha = 1$.
The physical basis for implicitly choosing $\alpha =1$ is unstated by Onsager,
but the consequence is that it is the entropy dissipation itself
(less the Rayleigh dissipation function) that is maximised
rather than some multiple thereof.
As such, this has the appearance of being a physical quantity.
Of course as a matter of logic Onsager could equally have chosen
$\alpha = -1$,
in which case he would have had a Principle of Minimum Dissipation.
Perhaps Onsager,
like other proponents of the Principle of Maximum Dissipation,
implicitly chose $\alpha =1 $ by analogy with
the Second Law of Equilibrium Thermodynamics.

Onsager's variational function upon which he based his Principle
of Maximum Dissipation will next be compared with Attard's
second entropy function,\cite{AttardII,NETDSM}
which amongst other things gives a particular value and physical interpretation
to the quantity $\alpha$.
The point that can be made here
is the arbitrary nature of Onsager's functional.
As such it cannot be the non-equilibrium thermodynamic potential,
because any ambiguity in such a potential would have
measurable physical consequences.
Onsager's implicit choice, $\alpha=1$,
corresponds to maximising the entropy dissipation,
is no more justified than the choice $\alpha=-1$,
which would correspond to minimising the entropy dissipation;
both choices are in fact incorrect, as will now be shown.
Finally, despite the assertion
made by Onsager and quoted above,
no evidence is offered by him
that the variational function is necessary for the reciprocal relations.
(By design, the variational functional is sufficient
to yield the linear transport laws.)

A variety of variational principles
can be constructed from different combinations of
the quadratic dissipation functions
$\dot S({\bf J},{\bf X}) \equiv {\bf J} \cdot{\bf X}$,
$\phi({\bf J}) \equiv {\bf J} \cdot \underline{\underline A}^{-1}  {\bf J} /2$,
and
$\psi({\bf X}) \equiv {\bf X} \cdot \underline{\underline A}  {\bf X} /2$.
These have featured in a number of postulated thermodynamic potentials
for non-equilibrium systems in the literature,
as has been reviewed by the present author.\cite{Attard08}
By design these always give the linear transport laws as
their optimum state,
but they are subject to the same criticisms
as have been made above:
In general the functions have no physical meaning
away from the optimum state,
and so they don't provide the actual driving force
toward the optimum state.
Also, there is no physical basis for extremising the dissipation,
and therefore there is no fundamental thermodynamic justification
for the postulated functions.

%%%%%%%%%%%%%%%%%%%%%%%%%%%%%%%%%%%%%%%%%%%%%%%%%%%%%%%%%%%%%%%%%%%%%%%%%
\subsection{The Second Entropy}

The present author gave a variational principle for the fluxes and structure
based upon the so-called second entropy,
which can also be called the transition entropy, or the two-time entropy.
The full theory is given elsewhere.\cite{AttardII,NETDSM}
Here it suffices to say that
the theory treats as the fundamental non-equilibrium object
the transition between two states
in a specified time interval,
${\bf x} \stackrel{\tau}{\rightarrow} {\bf x}'$.
The flux is essentially defined as the coarse derivative,
${\bf J} \equiv [{\bf x}'-{\bf x}]/\tau V$.
In fluctuation form,
the first (or ordinary, or structural)  entropy is
\begin{equation}
S^{(1)}({\bf x}) = \frac{1}{2}
{\bf x} \cdot \underline{ \underline S} {\bf x} ,
\end{equation}
the thermodynamic force is
${\bf X} = \underline{ \underline S} {\bf x}$,
and the second entropy is,
(in the present notation),
\begin{eqnarray} \label{Eq:S2}
\frac{ S^{(2)}({\bf J},{\bf x}) }{ V }
& = &
\frac{-|\tau| }{4} {\bf J} \cdot \underline{ \underline A}^{-1} {\bf J}
+ \frac{\tau}{2} {\bf J} \cdot {\bf X}
\nonumber \\ && \mbox{ }
+ \frac{|\tau|-2\tau}{4} {\bf X} \cdot \underline{ \underline A} {\bf X}
+ \frac{ S^{(1)}({\bf x}) }{ V }.
\end{eqnarray}
This holds for arbitrary fluxes that do not necessarily obey the linear
transport laws.
The all-important coefficient of the cross term,
which makes it half the dissipation,
as well as the final two terms,
which are constant with respect to the fluxes
and which are therefore of lesser importance,
come from a small time expansion
and the so-called reduction condition,
\begin{equation}
S^{(2)}(\overline {\bf J},{\bf x}) = S^{(1)}({\bf x}) .
\end{equation}
This is formally exact
and is simply a statement of Boltzmann's
molecular definition of entropy.\cite{AttardII,NETDSM}

It is evident that the second entropy is maximised
with respect to the fluxes
when the linear transport laws are satisfied,
\begin{equation}
\left.
\frac{ \partial S^{(2)}({\bf J},{\bf x})}{ \partial {\bf J} }
\right|_{ {\bf J} =\overline {\bf J} }
= {\bf 0}
\; \Leftrightarrow \;
\overline {\bf J}
=
\underline{ \underline A} {\bf X} ,
\end{equation}
going forward in time,
$\tau > 0$.
Hence maximising the second entropy
is equivalent to the linear transport laws.
But the second entropy itself is valid for
non-optimal fluxes that do not satisfy the linear transport laws.

From the nature of the derivation of the second entropy,
specifically that the fluctuations in an equilibrium system
are time symmetric,
the coefficient matrix is symmetric,
$\underline{ \underline A} = \underline{ \underline A}^\mathrm{T}$
(at least for variables of pure time parity).
Hence the second entropy implies the Onsager reciprocal relations.

%%%%%%%%%%%%%%%%%%%%%%%%%%%%%%%%%%%%%%%%%%%%%%%%%%%%%%%%
\subsubsection{Comparison with Onsager's First Function}

Comparing Onsager's variational function Eq.~(\ref{Eq:Onsager-vary})
with Attard's second entropy, Eq.~(\ref{Eq:S2})
going forward in time, $\tau > 0$,
one sees that apart from the immaterial terms
that are constant with respect to the fluxes,
the two are proportional to each other
\begin{equation}
{\cal O}({\bf J}|{\bf X})
=
\frac{2\alpha}{\tau} S^{(2)}({\bf J},{\bf X})
+ \mbox{const}.
\end{equation}
This says that the physically correct value is $\alpha = \tau/2$
(at least going forward in time).
As mentioned above,
any variational principle for the non-equilibrium thermodynamic potential
must be a quadratic in the fluxes,
and so the only choice in the functional form is for the three coefficients
of the quadratic equation.
Since only differences in potentials have physical meaning,
any constant term is immaterial,
and so there are two coefficients to determine.
One coefficient is fixed by demanding that the optimum flux
satisfy the linear transport laws.
The remaining degree of freedom
was fixed by Onsager by demanding that it be the rate of entropy dissipation
itself that be maximised, $\alpha = 1$.
There is no physical requirement for this.
In Attard's second entropy case,
this remaining degree of freedom was instead fixed
by the formally exact requirement
that the second entropy reduce to the first entropy
in the optimum state.
(The analysis was actually a little more complicated than this
in that it required a small time expansion of the fluctuation matrices,
equating the coefficients term by term in the reduction condition.)%
\cite{AttardII,NETDSM}

As mentioned above,
the arbitrary nature of the justification for
Onsager's functional actually has physical consequences
that preclude it from being a proper thermodynamic potential.
Onsager's implicit choice of $\alpha = 1$
means that his function ${\cal O}({\bf J}|{\bf X})$
and the second entropy $S^{(2)}({\bf J},{\bf X})$
have different curvatures.
If the functionals  are to represent the
non-equilibrium thermodynamic potential,
then their exponential must give the probability distribution
of the fluxes,
and their curvature must give the fluctuations in the fluxes.
Because the second entropy is the physical entropy,
it gives the probability
of a fluctuation in fluxes and structure
in an equilibrium system.
The arbitrary choice $\alpha=1$ by Onsager
yields incorrect fluctuations in the fluxes.

%%%%%%%%%%%%%%%%%%%%%%%%%%%%%%%%%%%%%%%%%%%%%%%%%%%%%%%%%
\subsubsection{Comparison with Onsager's Second Function}

Onsager, in Eq.~(5.10) of his second paper
on the reciprocal relations,\cite{Onsager31b}
gives what he believes is the actual thermodynamic potential whose
exponential divided by Boltzmann's constant
gives the probability of a transition.
It ought to be directly comparable to the second entropy,
and in the present notation it is
\begin{eqnarray}
\tilde{\cal O}({\bf J},{\bf x})
& = &
S^{(1)}({\bf x}) + S^{(1)}({\bf x}')
- \frac{\tau}{2} {\bf J} \cdot \underline{ \underline A}^{-1} {\bf J}
\nonumber \\ & = &
\mbox{const.} +
\tau {\bf J} \cdot {\bf X}
- \frac{\tau}{2} {\bf J} \cdot \underline{ \underline A}^{-1} {\bf J} .
\end{eqnarray}
Here an expansion to linear order in the time interval has been performed.
It can be seen that this differs by a factor of 2
from the second entropy.
In addition it does not correctly take into account
the irreversibility of thermodynamic transitions
(i.e.\ the absolute value of the time interval should appear
in the final term).
These discrepancies arise from the lack of physical justification
for Onsager's postulate that the rate of entropy dissipation is maximised.
Despite these problems,
one can nevertheless observe that this variational function is
rather close in spirit to the second entropy approach.

%%%%%%%%%%%%%%%%%%%%%%%%%%%%%%%%%%%%%%%%%%%%%%%%%%%%%%%%
\subsubsection{Comparison with Onsager's Third Function}

Onsager and Machlup,\cite{Onsager53}
some 20 years after the papers just mentioned,
gave a further variational principle
that is based upon an expression almost identical
to the second entropy expression.
In Eq.~(4.2),
they derive the conditional probability
for the scalar transition $x_1 \stackrel{\tau}{\rightarrow} x_2$
from the Langevin equation.
The key assumptions, in the present notation,
are that the first entropy has fluctuation form
(i.e.\ $S^{(1)}(x) = S x^2/2$ and $X = Sx$),
that the system is Markovian so that the most likely regression is exponential,
$\overline x_2 = e^{|\tau| V A S} x_1$,
and that the stochastic process is stationary and Gaussian.
The latter assumption means that the variance
of the random Langevin force is fully determined
by the entropy and transport constants.
This supplies the additional condition needed to fully determine
the two non-trivial coefficients for the quadratic variational principle
and removes the ambiguity of Onsager's prior versions of the Principle.

It is worth mentioning that a full analysis of
stationary, Gaussian, Markov processes,
which are also called Ornstein-Uhlenbeck processes,
is given by Keizer in \S 1.8 of Ref.~\onlinecite{Keizer87}.
Keizer treats the multi-component Langevin equation,
taking into account the non-commutativity of the matrices,
and gives the generalised fluctuation-dissipation theorem.
The generalised Langevin equation %(Markov and non-Markov)
is treated by the present author
in Ch.~10 of Ref.~\onlinecite{NETDSM}.

The second entropy corresponding to
the exponent of the unconditional probability
based on Eq.~(4.2)\cite{Onsager53}
is, in the present notation,
\begin{eqnarray}
\lefteqn{
S^{(2)}_\mathrm{OM}({x_2},{x_1}|\tau)
} \nonumber \\
& = &
%\frac{1}{2} \underline{\underline S}:{\bf x_1}{\bf x_1}
S^{(1)}({x_1})
%\nonumber \\ && \mbox{ }
+
\frac{1}{2} {S}
\left[ 1  -
e^{ 2 \tau V { A} { S} } \right]^{-1}
\left[ { x_2} - e^{ \tau  V { A} { S} } {x_1} \right]^2
%%%%%%%%%%%%%%%%%%%%%%%%%%%%%%%%%%%%%%%
 \nonumber \\ & = &
S^{(1)}({x_1})
-
\frac{1}{4 V \tau} {A}^{-1}
\left[ 1 + {\cal O}(\tau)  \right]
\nonumber \\ && \mbox{ } \times
\left[ {x_2} - {x_1} - \tau V {A}  S {x_1} + {\cal O}(\tau^2) \right]^2
%%%%%%%%%%%%%%%%%%%%%%%%%%%%%%%%%%%%%%%
 \nonumber \\ & = &
S^{(1)}({x_1})
-
\frac{\tau V}{4 } {A}^{-1}
%\nonumber \\ && \mbox{ }
\left[  {J}
- {A} {X_1}
\right]^2 + {\cal O}(\tau^2)  .
%%%%%%%%%%%%%%%%%%%%%%%%%%%%%%%%%%%%%%%
%\nonumber \\ & = &
%S^{(1)}({x_1})
%- \frac{\tau  V }{4 } { A}^{-1} { J}^2
%+ \frac{\tau V }{2 } { J}  {X_1}
%- \frac{ \tau V }{4 } { A}  {X}_1^2
%\nonumber \\ && \mbox{ }
%+ {\cal O}(\tau^2)
\end{eqnarray}
For $\tau > 0 $,
this may be seen to equal the expression for the second entropy
given above, Eq.~(\ref{Eq:S2}).
(For $\tau < 0 $ this does not correctly account for the irreversibility
of the trajectory.
Also, whereas Onsager assumed Markov behavior,
Eq.~(\ref{Eq:S2}) has been shown to hold as well for non-Markov systems.
See \S2.3 and \S  2.4.6 of Ref.~\onlinecite{NETDSM}.)

Onsager and Machlup,
in Eq.~(4.19) of Ref.~\onlinecite{Onsager53},
recast $S^{(2)}_\mathrm{OM}$
as  a variational principle
for the time integral of the so-called thermodynamic Lagrangian,
which involves the dissipation functions
$\phi$, $\psi$, and $\dot S$,
\begin{eqnarray}
{\cal O}_3({x}_2|{x}_1,\tau)
&=&
\frac{-1}{2} \left\{ \int_{t_1}^{t_2} \mathrm{d}t\,
\left[ \phi(\dot{x}(t))
+ \psi({X}({x}(t)))
\right. \right. \nonumber \\ && \left. \left. \mbox{ }
- \dot S(\dot{x}(t),{X}({x}(t)))
\right] \right\}_\mathrm{min} .
\end{eqnarray}
Minimisation of the integral
with respect to the path constrained to fixed values at specified nodes
(in this case $x_1$ and $x_2$ at $t_1$ and $t_2 = t_1+\tau$)
yields $S^{(2)}_\mathrm{OM}({x_2},{x_1}|\tau)$.
(It is possible to specify values at more than two nodes.)
This variational functional is rather common in the field
of stochastic differential equations
and there are a number of thermodynamic
Lagrangians that are based upon it.
\cite{Haken76, Graham77, Grabert79, Hunt81, Lavenda85,
Keizer87, Eyink90, Peng95}
In particular,
the integrand is the negative of the variational principle used by
Gyarmati,\cite{Gyarmati68,Gyarmati70}
and it is equal to the thermodynamic Lagrangian given by Lavenda,
Eq.~(1.17).\cite{Lavenda85}
The same criticism may be made of it as of the other
variational functions:
it has no physical meaning in the constrained state,
and it is but one of an infinite family of
variational functionals that could be constructed
to yield the steady state,
but which are physically meaningless more generally.

In any case, it is not entirely clear why one would want to
carry out this variational procedure since one already
has an explicit expression for both the optimum path between
the nodal values and the path entropy for the nodal values.
Perhaps Onsager never fully appreciated the second entropy,
and instead continued to search
for an alternative thermodynamic variational principle
for non-equilibrium systems based upon the rate of entropy dissipation.
From the point of view of the present author,
the thermodynamic principle for non-equilibrium systems
is given by the second entropy for transitions,
and this is not directly connected to the rate of entropy dissipation.

%%%%%%%%%%%%%%%%%%%%%%%%%%%%%%%%%%%%%%%%%%%%%%%%%%%%%%%%%
\subsubsection{The Point of the Second Entropy}

What is the use of the second entropy? Does it give anything beyond
the already known reciprocal relations and linear transport laws?
One can identify two additional results of significance. First, and
conceptually, it provides a rigorous  variational principle for
non-equilibrium thermodynamics, and one that is unique on physical
grounds. This is the analogue of the Second Law of Equilibrium
Thermodynamics. Second, and practically, it provides the correct
formula for the probability of fluctuations in the fluxes. These are
not given by the linear transport laws, and it is this property that
makes it the correct non-equilibrium thermodynamic potential.

Beyond these specific consequences,
there is more nebulous but possibly more far-reaching outcome
of the second entropy approach:
it focusses thinking about non-equilibrium systems
on the transitions between states rather than on the states themselves.
The states may be regarded as the objects of equilibrium thermodynamics
and, in contrast, it is the transitions
that are the stuff of non-equilibrium thermodynamics.
It is this change in mind set that is crucial to providing
the insight as to why the Principle of Least Dissipation
(or Principle of Maximum Dissipation)
cannot provide the basis for non-equilibrium thermodynamics.
And, as will be shown explicitly in this paper,
it provides a conceptual basis for
interpreting not only the specific computational results
obtained here for heat convection,
but also for understanding in general
non-equilibrium phase transitions
and pattern formation in non-equilibrium systems.

%%%%%%%%%%%%%%%%%%%%%%%%%%%%%%%%%%%%%%%%%%%%%%%%%%%%%%%%%%%%%%%%%%
%
\section{Hydrodynamic Equations of Convection} \label{Sec:convheatflow}
%
%%%%%%%%%%%%%%%%%%%%%%%%%%%%%%%%%%%%%%%%%%%%%%%%%%%%%%%%%%%%%%%%%%

%%%%%%%%%%%%%%%%%%%%%%%%%%%%%%%%%%%%%%%%%%%%%%%%%
\subsection{Boussinesq Approximation}

The Boussinesq approximation
is generally invoked for
hydrodynamic calculations of convective heat flow.\cite{Yih77,Drazin81}
In this the compressibility is set to zero, $\chi_\mathrm{T} = 0$,
and the thermal expansivity is neglected everywhere
except in the buoyancy force.
With these the density equation reduces to the vanishing
of the divergence of the velocity field,
\begin{equation}
\nabla \cdot {\bf v}({\bf r},t) = 0 .
\end{equation}
This means that the most likely value of the scalar part of the
viscous pressure tensor vanishes, $\overline \pi = 0$.

The gravitational potential density is
\begin{equation}
n({\bf r},t) \psi({\bf r},t) =
\left\{ n_{00} - \alpha n_{00} [T_\mathrm{tot}({\bf r},t) - T_{00} ]
\right\} m g z .
\end{equation}
Here
$\alpha$ is the thermal expansivity,
$g$ is the acceleration due to gravity, and $m$ is the molecular mass.
The subscript tot signifies the total temperature,
$T_\mathrm{tot} = T_0 + T$,
where $T_0$ is the temperature in conduction,
and $T$ is convective perturbation.
The subscript 00 denotes the reference value at the
mid-point of the sub-system,
and everything will be expanded to linear order in the
difference from this reference point.
Inserting this in the Navier-Stokes equation
and neglecting the term quadratic in the velocity yields
\begin{eqnarray}
m n_{00}  \frac{\partial  {\bf v}({\bf r},t)}{\partial t}
& = &
- \left\{ n_{00} - \alpha n_{00} [ T_\mathrm{tot}({\bf r},t) - T_{00} ] \right\}
m g \hat{\bf z}
\nonumber \\ && \mbox{ }
- \nabla p_\mathrm{tot}({\bf r},t)
+ \eta \nabla^2 {\bf v}({\bf r},t) ,
\end{eqnarray}
where $\eta$ is the shear viscosity.

Neglecting the viscous dissipation, which is quadratic in the
velocity, and also the thermal expansivity, and using the most
likely heat flux, the energy equation becomes
\begin{equation}
c_\mathrm{p} \frac{\partial T_\mathrm{tot}({\bf r},t)}{\partial  t}
+
c_\mathrm{p} {\bf v}({\bf r},t) \cdot \nabla T_\mathrm{tot}({\bf r},t)
= \lambda \nabla^2 T_\mathrm{tot}({\bf r},t) ,
\end{equation}
where $\lambda$ is the thermal conductivity.
The coupling of velocity and temperature represents a non-linear term.
These three partial differential equations constitute the Boussinesq
approximation that is to be solved for the temperature, pressure,
and velocity fields.

%%%%%%%%%%%%%%%%%%%%%%%%%%%%%%%%%%%%%%%%%%%%%%%%%
\subsection{Conduction}

The simplest case of slab geometry is treated here,
with a temperature gradient imposed in the $z$-direction.
The convective flow is treated as a perturbation
from the conducting state.
In conduction, the velocity field is zero, ${\bf v}({\bf r},t) = 0$.

The boundaries of the sub-system are located at $z = \pm L_z/2$, and
the temperatures of the reservoirs beyond the boundaries are
$T_\mathrm{r\pm}$.
The temperature difference is
$\Delta_\mathrm{T} \equiv T_\mathrm{r+} - T_\mathrm{r-} $.
It is assumed that the
imposed temperature gradient, $\Delta_\mathrm{T}/L_z  $, is small
and that quadratic terms can be neglected.
This means that it does
not matter whether one deals with the difference in temperature or
with the difference in inverse temperature. For convection to occur,
the lower reservoir must be hotter than the upper reservoir,
$\Delta_\mathrm{T} < 0 $.

Since in conduction the velocity vanishes, and the temperature is steady and a
function of $z$ only, the energy equation reduces to
\begin{equation}
0 =\lambda  \frac{\mathrm{d}^2 T_0(z)}{\mathrm{d} z^2}  .
\end{equation}
Hence the temperature field is
a linear function of $z$ that must equal the reservoirs'
temperatures at the boundaries,
\begin{equation}
T_0(z) = T_{00} + \frac{ \Delta_\mathrm{T} }{ L_z } z
, \; |z| \le L_z/2 ,
\end{equation}
with the mid-point temperature being
$T_{00} \equiv [T_\mathrm{r+} + T_\mathrm{r-}]/2$.
With this and zero velocity, the Navier-Stokes
equation becomes
\begin{equation}
0 =
- \left\{ n_{00} - \frac{ \alpha n_{00} \Delta_\mathrm{T} }{ L_z } z
\right\} m g
-  \frac{\mathrm{d} p_0(z)}{\mathrm{d} z} .
\end{equation}
Hence the  pressure profile in conduction is
\begin{equation}
p_0(z) = p_{00} - n_{00} m g z
+  \frac{ \alpha n_{00} m g \Delta_\mathrm{T} }{ 2 L_z } z^2 .
\end{equation}

For future use, the heat flow  per unit area in conduction is
\begin{equation}
\overline{ J_\mathrm{E,0}^0 }
= - \lambda \frac{\mathrm{d} T_0(z)}{\mathrm{d}  z}
= \frac{ - \lambda \Delta_\mathrm{T} }{ L_z } .
\end{equation}
The rate of entropy production per unit sub-system volume  is
\index{entropy, equilibrium!rate of change}%%%
\begin{equation}
\dot S_\mathrm{r} /AL_z
=
\frac{ \overline{ J_\mathrm{E,0}^0 }}{L_z}
\left[ \frac{1}{T_\mathrm{r+}} - \frac{1}{T_\mathrm{r-}} \right]
=  \frac{ \lambda \Delta_\mathrm{T}^2 }{ T_{00}^2 L_z^2 } .
\end{equation}
This is positive  and independent of the sign of the temperature
difference, as it ought to be.

%%%%%%%%%%%%%%%%%%%%%%%%%%%%%%%%%%%%%%%%%%%%%%%%%
\subsection{Convection} \label{Sec:NSconv}

Regarding convection as a perturbation on conduction,
the temperature may be written
\begin{equation} \label{Eq:TB}
T_\mathrm{tot}({\bf r},t) = T_0(z) + T({\bf r},t) ,
\end{equation}
and similarly  the pressure,
\begin{equation}
p_\mathrm{tot}({\bf r},t) = p_0(z) +  p({\bf r},t) .
\end{equation}
Since the velocity is zero in conduction,
the full velocity field is the same as the perturbing velocity field,
${\bf v}_\mathrm{tot}({\bf r},t) = {\bf v}({\bf r},t)$.
These convective fields depend upon the Rayleigh number
and, in the calculations below,
the wave number of the steady state being characterised,
but these will not be shown explicitly.

The full fields satisfy the density, Navier-Stokes, and energy equations.
But since the conductive fields
also satisfy these equations,
they can be subtracted from both sides, so that one has
\begin{equation}
0 = \nabla \cdot {\bf v}({\bf r}) ,
\end{equation}
\begin{equation}
m n_{00}  \frac{\partial  {\bf v}({\bf r},t)}{\partial t}
 =
\alpha n_{00} T({\bf r}) m g \hat{\bf z}
- \nabla p({\bf r})
%\nonumber \\ && \mbox{ }
+ \eta \nabla^2 {\bf v}({\bf r}) ,
\end{equation}
and
\begin{eqnarray}
c_\mathrm{p} \frac{\partial  T({\bf r},t)}{\partial  t}
& = &
- c_\mathrm{p} {\bf v}({\bf r}) \cdot \nabla [T_0(z)
+  T({\bf r})] + \lambda  \nabla^2 T({\bf r})
\nonumber \\ & = &
- \frac{ c_\mathrm{p} \Delta_\mathrm{T}}{L_z} v_z({\bf r})
- c_\mathrm{p} {\bf v}({\bf r}) \cdot \nabla T({\bf r})
+ \lambda  \nabla^2 T({\bf r}) .
\nonumber \\
\end{eqnarray}
In the steady state the left-hand sides are zero.
Here the left hand side of the final term will be retained
as non-zero.
This is useful both as an iterative device
(simple time stepping)
for the computer algorithm to obtain a converged
steady state solution,
and also to obtain the physical properties of the system
during the evolution in time either from conduction to convection
of during the transition from one convecting state to another.
This implicitly assumes rapid relaxation
of the velocity at each time step, ${\partial  {\bf v}}/{\partial t} = 0$.

If one regards the convective perturbation as small, then one sees
that the second term on the right-hand side of the energy equation
is non-linear, as it is the product of the convective temperature and
the velocity. This non-linear term fixes the amplitude of the fields
that give the steady state, since without it everything could be
multiplied by a constant to give another solution. There are five
equations (the Navier-Stokes equation is for three components) and
five fields, including the three components of the velocity.

Now use $L_z$ as the unit of length, $-\Delta_\mathrm{T}$ as the
unit of temperature, $L_z^2 c_\mathrm{p} /\lambda$ as the unit of
time, and $ m n_{00} \lambda^2 /L_z^2  c_\mathrm{p}^2$ as the unit
of pressure. Denoting dimensionless quantities with an asterisk, one
has
\begin{equation}
0 = \nabla^* \cdot {\bf v}^* ,
\end{equation}
\begin{equation}
\frac{\partial  {\bf v}^*}{\partial t^*}
= {\cal R}{\cal P}  T^* \hat{\bf z}
- \nabla^*  p^* + {\cal P} \nabla^{*2} {\bf v}^* ,
\end{equation}
and
\begin{equation}
\frac{\partial  T^*}{\partial  t^*}
= v_z^* -  {\bf v}^* \cdot \nabla^*  T^*
+  \nabla^{*2}  T^* .
\end{equation}
Here the Rayleigh number is
\index{Rayleigh number}%%%
\begin{equation}
{\cal R} \equiv
- \Delta_\mathrm{T} \alpha g c_\mathrm{p} m n_{00} L_z^3 /\lambda \eta ,
\end{equation}
and the Prandtl number is
\index{Prandtl number}%%%
\begin{equation}
{\cal P} \equiv \eta c_\mathrm{p} / m n_{00} \lambda  .
\end{equation}
Here and throughout, $c_\mathrm{p}$ is the constant pressure heat
capacity per unit volume.

One can eliminate the pressure from the Navier-Stokes equations. Set
the left-hand side to zero, differentiate the  $z$-component with
respect to $y$, the $y$-component with respect to $z$, and subtract,
\begin{equation}
0 =
{\cal R} \frac{ \partial  \tilde T^* }{ \partial y^*}
+
\nabla^{*2} \left[ \frac{ \partial v^*_z }{ \partial y^*}
- \frac{ \partial v^*_y }{ \partial z^*} \right] .
\end{equation}
The Prandtl number has been factored out. Similarly for the
$x$-component,
\begin{equation}
0 =
{\cal R} \frac{ \partial  \tilde T^* }{ \partial x^*}
+
\nabla^{*2} \left[ \frac{ \partial v^*_z }{ \partial x^*}
- \frac{ \partial v^*_x }{ \partial z^*} \right] .
\end{equation}
One now has four equations
(these two forms of the momentum equation,
the density equation, and the energy equation),
four fields ($T^*$, $v^*_x$, $v^*_z$, and $v^*_z$),
and one dimensionless parameter, ${\cal R}$.
Since everything in these equations  is dimensionless and refers to
the convective perturbation,
the asterisk will be dropped later below.

%%%%%%%%%%%%%%%%%%%%%%%%%%%%%%%%%%%%%%%%%%%%%%%%%%%%%%%%%%%%%%%%%%
%
\section{Total First Entropy of Convection} \label{Sec:s1}
%
%%%%%%%%%%%%%%%%%%%%%%%%%%%%%%%%%%%%%%%%%%%%%%%%%%%%%%%%%%%%%%%%%%

Now an expression for the first or structural
entropy of a convecting steady state will be obtained
as the difference from the conducting state.
(Quite generally the free energy is minus the temperature
times the total first entropy,\cite{TDSM}
and so the following results could be recast in terms
of free energy, if desired.)
The total entropy is the sum of the sub-system entropy
and the reservoir entropy.
Here the exact sub-system entropy will be given,
and two forms for the reservoir contribution will be obtained.
One reservoir expression  is the exact change in
reservoir entropy during the transition from one state
to another (e.g.\ conduction to convection,
or from one convecting state to another).
The second reservoir expression is the so-called static approximation.
It gives the difference in reservoir entropy between a convecting state
and the conducting state,
and again it can be used to obtain the difference in reservoir entropy
between one convecting state and another.

The static approximation for the reservoir entropy
in a non-equilibrium system was originally presented
in Ref.~\onlinecite{AttardV},
and its nature is discussed in full detail in Ch.~9 of Ref.~\onlinecite{NETDSM}.
Briefly,
the first entropy for phase space for a non-equilibrium system
consists of a static part,
which is the analogue of the usual equilibrium expression,
and a dynamic part, which is a correction to the static reservoir contribution
that accounts for the adiabatic evolution that is unavoidably included.
In the present macroscopic description,
the sub-system entropy is given exactly by the usual equilibrium expression,
which will be obtained explicitly.
The non-equilibrium reservoir entropy
will be approximated by neglecting the dynamic part
and using the static part alone.

The change (or difference) in entropy density %(per unit volume)
of the sub-system between convection and conduction
can be obtained by thermodynamic integration of the
temperature from that in conduction,
$T_0(z) = T_{00} + z \Delta_T /L_z $,
to that in convection
$T_\mathrm{tot}({\bf r}) = T_0(z) + T({\bf r})$.
(The dependence of the temperature field
(and of the change in entropy)
on the Rayleigh number and on the wave number of the particular
convective state is not shown explicitly.
Also, dimensionless variables are \emph{not} used in this section.)
Henceforth this will simply be called the convective entropy density,
the change from conduction being understood.
Assuming, as in the Boussinesq approximation,
that the thermal expansivity and compressibility can be neglected,
the convective entropy density of the sub-system is
\begin{eqnarray} \label{Eq:sigma-s}
\sigma_\mathrm{s}({\bf r}) & = &
\int_{\varepsilon_\mathrm{int,0}}^{\varepsilon_\mathrm{int,1}}
\mathrm{d} \varepsilon_\mathrm{int}'  \;
\frac{\partial \sigma(\varepsilon_\mathrm{int}')
}{ \partial \varepsilon_\mathrm{int}' }
\nonumber \\ & = &
\int_{\varepsilon_\mathrm{int,0}}^{\varepsilon_\mathrm{int,1}}
\mathrm{d} \varepsilon_\mathrm{int}'  \;
\frac{1}{T'}
\nonumber \\ & = &
c_\mathrm{p} \ln\left[ \frac{T_0(z)+ T({\bf r}) }{T_0(z)} \right]
\nonumber \\ & = &
c_\mathrm{p}  \left[ \frac{T({\bf r}) }{T_0(z)}
-  \frac{ T({\bf r})^2 }{2T_0(z)^2} +  \ldots \right]
\nonumber \\  & = &
c_\mathrm{p}  \left[ \frac{T({\bf r}) }{T_{00}}
- \frac{z \Delta_T T({\bf r})}{L_z T_{00}^2}
- \frac{T({\bf r})^2  }{2T_{00}^2}
+  {\cal O}\!\left(\Delta_T^3/T_{00}^3\right) \right] .
\nonumber \\
\end{eqnarray}
This uses the fact that
$\Delta \varepsilon_\mathrm{int} = c_\mathrm{p}  \Delta T$,
where $c_\mathrm{p}$ is the heat capacity per unit volume.
Since the convective temperature is
$ T({\bf r}) \sim {\cal O}(\Delta_T)$,
the expansion to quadratic order
in the temperature difference is justified.
(This expansion has been checked numerically against
the full logarithm and found to be accurate
for ideal straight rolls over the full range of
Rayleigh numbers and wave numbers.)
This is the local sub-system entropy density.
The global sub-system entropy density is
\begin{equation} \label{Eq:sigma-st}
\sigma_\mathrm{s}(t)
=
\frac{1}{AL_z} \int \mathrm{d}{\bf r}\;
\sigma_\mathrm{s}({\bf r},t) ,
\end{equation}
where $A$ is the cross-sectional area of the sub-system.
For use below during convective transitions,
this has been written as a function of time
and invokes the instantaneous temperature, $T({\bf r},t)$.

%%%%%%%%%%%%%%%%%%%%%%%%%%%%%%%%%%%%%%%%
\subsubsection{Static Reservoir Entropy}

The zeroth and first temperatures of the reservoirs are\cite{AttardI}
\begin{equation}
\frac{1}{T_{\mathrm{r}0}} =
\frac{1}{2}\left[
\frac{1}{T_{\mathrm{r}+}} + \frac{1}{T_{\mathrm{r}-}}
\right]
= \frac{1}{T_{00}} +{\cal O} \!\left(\Delta_T^2/T_{00}^2\right) ,
\end{equation}
and
\begin{equation}
\frac{1}{T_{\mathrm{r}1}} =
\frac{1}{L_z}\left[
\frac{1}{T_{\mathrm{r}+}} - \frac{1}{T_{\mathrm{r}-}}
\right]
= \frac{-\Delta_T}{L_z T_{00}^2 } + {\cal O}\!\left(\Delta_T^3/T_{00}^3\right),
\end{equation}
respectively.
These correspond in essence to the average temperature
and to the temperature gradient of the reservoirs.
With these, the static part of the reservoir entropy
associated with the sub-system is
\begin{equation}
S_\mathrm{r,st} =
\frac{-\Delta E_0}{T_{\mathrm{r}0}} - \frac{\Delta E_1}{T_{\mathrm{r}1}} .
\end{equation}
Again it is understood that this is the change from conduction.
The energy moments of the sub-system are defined as
\begin{equation}
\Delta E_n =
\int \mathrm{d}{\bf r} \; z^n \Delta \varepsilon({\bf r}) .
\end{equation}
In view of this one can define
the static convective reservoir entropy density,
%that is associated with the state of the sub-system,
\begin{equation}
\sigma_\mathrm{r,st}({\bf r}) \equiv
\frac{-\Delta \varepsilon({\bf r}) }{T_{\mathrm{r}0}}
- \frac{z \Delta \varepsilon({\bf r}) }{T_{\mathrm{r}1}} .
\end{equation}

What appears here is the change in total energy density,
which is composed of the internal energy density,
the gravitational energy density, and the kinetic energy density,
\begin{eqnarray}
\Delta \varepsilon ({\bf r}) & = &
\Delta \varepsilon_\mathrm{int}({\bf r})
+ \Delta \varepsilon_\mathrm{g}({\bf r})
+ \Delta \varepsilon_\mathrm{ke}({\bf r})
\nonumber \\  & = &
c_\mathrm{p} T({\bf r})
- \alpha m n_{00} g z T({\bf r})
+ \frac{ m n_{00} }{2} {\bf v}({\bf r}) \cdot {\bf v}({\bf r}) .
\nonumber \\
\end{eqnarray}
Accordingly, the change in total entropy density,
$\sigma_\mathrm{tot,st}({\bf r})
=
\sigma_\mathrm{s}({\bf r}) + \sigma_\mathrm{r,st}({\bf r})$,
is composed of three analogous terms.
The internal energy density contribution
includes the sub-system entropy and is
\begin{eqnarray} \label{Eq:s-st-int}
\sigma_\mathrm{tot,st}^\mathrm{int}({\bf r})
& \equiv &
\sigma_\mathrm{s}({\bf r})
- \frac{\Delta \varepsilon_\mathrm{int}({\bf r})}{T_{\mathrm{r}0}}
- \frac{z \Delta \varepsilon_\mathrm{int}({\bf r})}{T_{\mathrm{r}1}}
\nonumber \\  & = &
c_\mathrm{p}  \left[ \frac{T({\bf r}) }{T_{00}}
- \frac{z \Delta_T T({\bf r})}{L_z T_{00}^2}
- \frac{T({\bf r})^2  }{2T_{00}^2} \right]
\nonumber \\  & & \mbox{ }
- \frac{c_\mathrm{p} T({\bf r})}{T_{00}}
+ \frac{c_\mathrm{p} z \Delta_T T({\bf r})}{L_z T_{00}^2 }
\nonumber \\  & = &
\frac{-  c_\mathrm{p}  }{2T_{00}^2}  T({\bf r})^2.
\end{eqnarray}
This is identical to the equilibrium fluctuation expression
for the total entropy density of a sub-system in equilibrium with a reservoir
of temperature $T_{00}$ when the local fluctuation in energy is
$ \Delta \varepsilon_\mathrm{int}({\bf r}) =  c_\mathrm{p} T({\bf r})$.
It is what one would have expected and could have been written down directly.
The fact that this is always negative
means that a convecting steady state
has lower total entropy than the conducting state,
at least as far as the rearrangement of the internal energy
in the convecting system is concerned.
This latter observation is not particularly significant
because the result depends upon the static approximation,
and so it does not give the full change in the reservoir entropy
during such a transition.

The gravitational contribution is
\begin{eqnarray} \label{Eq:s-st-g}
\sigma_\mathrm{tot,st}^\mathrm{g}({\bf r})
& \equiv &
- \frac{\Delta \varepsilon_\mathrm{g}({\bf r})}{T_{\mathrm{r}0}}
- \frac{z \Delta \varepsilon_\mathrm{g}({\bf r})}{T_{\mathrm{r}1}}
\\ \nonumber  & = &
\frac{\alpha m n_{00} g z T({\bf r})}{T_{00}}
- \frac{\alpha m n_{00} g z^2 \Delta_T T({\bf r})}{L_z T_{00}^2 } ,
\end{eqnarray}
and the kinetic energy contribution is
\begin{eqnarray} \label{Eq:s-st-ke}
\sigma_\mathrm{tot,st}^\mathrm{ke}({\bf r})
& \equiv &
- \frac{\Delta \varepsilon_\mathrm{ke}({\bf r})}{T_{\mathrm{r}0}}
- \frac{z \Delta \varepsilon_\mathrm{ke}({\bf r})}{T_{\mathrm{r}1}}
\\ \nonumber  & = &
- \frac{ m n_{00} }{2T_{00}} {\bf v}({\bf r}) \cdot {\bf v}({\bf r})
+
\frac{ m n_{00} \Delta_T z }{2L_z T_{00}^2 }
{\bf v}({\bf r}) \cdot {\bf v}({\bf r}) .
\end{eqnarray}
In the case of ideal straight rolls,
there is an up-down symmetry
so that the convective temperature perturbation
is anti-symmetric upon reflection through the center
of a convective roll.
This means that the final term on the right-hand side of each of these
integrates to zero.

The sum of the last three results
represent the static approximation to the convective entropy density
(i.e.\ the difference in entropy between the convective state
and the conductive state).
Integrating over the volume of the sub-system,
the global convective entropy density is
\begin{equation} \label{Eq:stot-st}
\sigma_\mathrm{tot,st}
\equiv
\frac{1}{AL_z} \int \mathrm{d}{\bf r} \;
\left[ \sigma_\mathrm{tot,st}^\mathrm{int}({\bf r})
+ \sigma_\mathrm{tot,st}^\mathrm{g}({\bf r})
+ \sigma_\mathrm{tot,st}^\mathrm{ke}({\bf r})
\right] .
\end{equation}

%%%%%%%%%%%%%%%%%%%%%%%%%%%%%%%%%%%%%%%%%%%
\subsubsection{Change in Reservoir Entropy}

The reservoir contribution to the above result for the convection entropy
is approximate, whereas the sub-system entropy is exact.
However, by integrating over time the heat flow
through the sub-system from one reservoir to the other,
one can obtain exactly the change in reservoir entropy
for a transition between two non-equilibrium states.

The formally exact rate of change of the entropy of the reservoirs is
\begin{eqnarray} \label{Eq:dSrt/dt}
\dot S_\mathrm{r}(t)
& = &
\int_A \mathrm{d}x \, \mathrm{d}y \,
\left[
\frac{1}{T_\mathrm{r+}} \overline J_\mathrm{E}^0(x,y,L_z/2,t)
\right. \nonumber \\ & & \left. \mbox{ }
-
\frac{1}{T_\mathrm{r-}} \overline  J_\mathrm{E}^0(x,y,-L_z/2,t)
\right]
\nonumber \\ & \approx &
\frac{-L_z \Delta_T}{T_{00}^2}
\int_A \mathrm{d}x \, \mathrm{d}y \,
\overline J_\mathrm{E}^0(x,y,L_z/2,t)
\nonumber \\ & = &
\frac{-L_z \Delta_T \lambda }{T_{00}^2}
\int_A \mathrm{d}x \, \mathrm{d}y \,
\left. \frac{\partial T({\bf r},t) }{\partial z} \right|_{z=L_z/2} .
\end{eqnarray}
The first equality is exact,
whereas the second equality makes the approximation
that the integrated heat flux at the two boundaries are equal.
This is certainly the case in the steady state,
and it is a very good approximation in the transitions
between steady straight roll states that are characterised below.
This approximation does not account for any nett
total energy change of the sub-system during a transition,
such as those in the gravitational energy
and in the kinetic energy, but these are negligible
compared to the total heat flux over the time interval of a transition.
Using this,
the change in total entropy per unit sub-system volume
during a transition over the time interval $ [t_1 , t_2 ]$ is
\begin{equation} \label{Eq:Stot-trans}
\Delta \sigma_\mathrm{tot}
=
\sigma_\mathrm{s}(t_2) - \sigma_\mathrm{s}(t_1)
+
\frac{1}{AL_z} \int_{t_1}^{t_2}  \mathrm{d}t \; \dot S_\mathrm{r}(t) ,
\end{equation}
where the global sub-system entropy density is given by Eq.~(\ref{Eq:sigma-st}).

%%%%%%%%%%%%%%%%%%%%%%%%%%%%%%%%%%%%%%%%%%%%%%%%%%%%%%%%%%%%%%%%%%%%%%%%%%
%
\section{Ideal Straight Rolls} \label{Sec:algo-strt}
%\setcounter{equation}{0} \setcounter{subsubsection}{0}
%
%%%%%%%%%%%%%%%%%%%%%%%%%%%%%%%%%%%%%%%%%%%%%%%%%%%%%%%%%%%%%%%%%%%%%%%%%%

This and the following section
set out the hydrodynamic equations used for convection
and describes the computer algorithms that were used to solve them.
This section deals with ideal straight rolls
(i.e.\ the rolls are considered straight and homogeneous in the
$x$-direction),
and the next section deals with the cross roll state
(i.e.\ the combination of straight $x$- and $y$-rolls).
%The latter is relevant to the so-called cross roll transition,
%which is the transition from straight rolls aligned with the  $x$-axis
%to straight rolls aligned with the $y$-axis.

The applied thermal gradient and gravity are in the $z$-direction.
The wavelength is twice the width of an individual
roll, $\Lambda = 2 L_y$, as they come in pairs of counter-rotating rolls,
and the wave number is defined as $a = 2 \pi/\Lambda$.

In this and the following sections,
dimensionless variables are used,
with the asterisk being dropped.
Hence $L_y = 1$ or $a \approx 3.1$ corresponds to a roll
whose width and height are approximately equal.

%%%%%%%%%%%%%%%%%%%%%%%%%%%%%%%%%%%%%%%%%%%%%%%%%%%%%%%%%%%%%%%%%%%%%%%%%%
\subsection{Hydrodynamic Equations}

The hydrodynamic equations for convection were given at the end of
\S\ref{Sec:NSconv}.
For the present ideal straight rolls
with their axis in the $x$-direction,
the $x$-component of velocity and the
$x$-derivatives are zero.
The three equations for the remaining three fields are
\begin{equation}
0 =
\frac{\partial v_y(y,z)}{\partial y}
+ \frac{\partial v_z(y,z)}{\partial z} ,
\end{equation}
\index{energy!equation@equation ${\mathrm{d}T({\bf r},t)}/{\mathrm{d}t}$}%%%
\begin{eqnarray} \label{Eq:dTdt1}
%\frac{\partial T}{\partial t}
\frac{\partial T(y,z)}{\partial t}
& = &
v_z(y,z)
- v_y(y,z) \frac{\partial T(y,z)}{\partial y}
- v_z(y,z) \frac{\partial T(y,z)}{\partial z}
\nonumber \\ &  & \mbox{ }
+ \frac{\partial^2 T(y,z) }{\partial y^2}
+ \frac{\partial^2  T(y,z)}{\partial z^2}  ,
\end{eqnarray}
and
\begin{eqnarray} \label{Eq:dpdt}
0 & = &
{\cal R} \frac{ \partial  T(y,z) }{ \partial y}
+ \nabla^{2} \left[ \frac{ \partial v_z(y,z) }{ \partial y}
- \frac{ \partial v_y(y,z) }{ \partial z} \right]
\nonumber \\ & = &
{\cal R} \frac{ \partial^2  T(y,z) }{ \partial y^2}
+ \left[ \frac{ \partial^2 }{ \partial y^2}
+ \frac{ \partial^2 }{ \partial z^2} \right]^2 v_z(y,z) .
\end{eqnarray}
The second equality follows by taking the $y$-derivative of the
first equality and using the vanishing of the divergence of the
velocity. Recall that the Rayleigh number is ${\cal R} \equiv
-\alpha m g n_{00}  c_\mathrm{p} \Delta_T L_z^3 /\lambda \eta$.

%%%%%%%%%%%%%%%%%%%%%%%%%%%%%%%%%%%%%%%%%%%%%%%%%%%%%%%%%%%%%%%%%%%%%%%%
\subsection{Fourier Expansion}

Following Busse,\cite{Busse67}
a Galerkin method is used that invokes Fourier expansions of the
fields. The temperature field is expanded as
\begin{eqnarray}
T(y,z) & = &
\sum_{l=0}^{L} \sum_{n=1}^{N} \left[ T_{ln}^\mathrm{s}
\sin 2 n\pi z
\right. \nonumber \\ && \mbox{ } \left.
+ T_{ln}^\mathrm{c} \cos( 2 n-1) \pi z \right] \cos l a y .
\end{eqnarray}
The form of the $z$-expansion
is chosen to guarantee the boundary conditions, $T(y,\pm 1/2) = 0$.
For the Boussinesq fluid, there is mirror plane symmetry between two
rolls, $T(y,z) = T(-y,z)$, and point reflection symmetry within a
roll, $T(y,z) = -T(L_y-y,-z)$. These mean that the even $l$
coefficients of $T_{ln}^\mathrm{c}$ and the odd $l$ coefficients of
$T_{ln}^\mathrm{s}$ must vanish.

The particular solution of the differential equation for the
velocity is
\begin{eqnarray}
v_z^\mathrm{p}(y,z) & = &
\sum_{l,n} \left[ v_{z,ln}^\mathrm{ps}
\sin 2 n\pi z
\right. \nonumber \\ && \mbox{ } \left.
+ v_{z,ln}^\mathrm{pc}  \cos( 2 n-1) \pi z \right] \cos l a y .
\end{eqnarray}
Clearly,
\begin{equation}
v_{z,ln}^\mathrm{ps} =
\frac{{\cal R}(l a )^2} {\left[ (l a )^2+(2 n\pi)^2\right]^2}
T_{ln}^\mathrm{s} ,
\end{equation}
and
\begin{equation}
v_{z,ln}^\mathrm{pc} =
\frac{{\cal R}(l a )^2}{\left[ (l a )^2+((2n-1)\pi)^2\right]^2}
T_{ln}^\mathrm{c} .
\end{equation}
%Obviously these depend upon the temperature coefficients
%at the current time.

The homogeneous solution, which satisfies $\nabla^2 \nabla^2
v_z^\mathrm{h} = 0$, is
\begin{eqnarray}
v_z^\mathrm{h}(y,z)
& = & \sum_{l=1}^L \left[ A_l^\mathrm{s}
\sinh l a z  + B_l^\mathrm{s} z \cosh  l a z
\right. \\ \nonumber && \mbox{ } \left.
+ A_l^\mathrm{c}  \cosh l a z  + B_l^\mathrm{c} z
\sinh  l a z \right] \cos l a y .
\end{eqnarray}
Because the system is periodic in the horizontal direction, there is
a term for each expansion mode. Writing the velocity as
$v_z = v_z^\mathrm{p} + v_z^\mathrm{h}$,
the four boundary conditions for each mode,
$v_z(y,\pm 1/2) = \partial v_z(y,\pm 1/2)/\partial z = 0$,
determine the four coefficients per mode, $ A_l^\mathrm{s}$,
$B_l^\mathrm{s}$ $ A_l^\mathrm{c} $, and  $B_l^\mathrm{c}$.
These coefficients have the Boussinesq symmetry discussed above. The
second condition ensures the vanishing of $v_y(y,\pm 1/2)$ when the
density equation is applied.

The vertical velocity field is then projected onto the Fourier grid
used for the temperature field using the orthogonality of the
trigonometric functions. Formally one has
\begin{eqnarray} \label{Eq:vzB}
v_z(y,z) & = &
\sum_{l=1}^{L} \sum_{n=1}^{N} \left[
v_{z,ln}^\mathrm{s}  \sin 2 n\pi z
\right. \nonumber \\ && \mbox{ } \left.
+ v_{z,ln}^\mathrm{c} \cos( 2 n-1) \pi z \right] \cos l a y .
\end{eqnarray}
Due to the Boussinesq symmetry, half the coefficients are zero. The
horizontal velocity may be expanded as
\begin{eqnarray} \label{Eq:vyB}
v_y(y,z) & = &
\sum_{l=1}^{L} \sum_{n=1}^{N} \left[
v_{y,ln}^\mathrm{c}  \cos 2 n\pi z
\right. \nonumber \\ && \mbox{ } \left.
+ v_{y,ln}^\mathrm{s} \sin( 2 n-1) \pi z \right] \sin l a y .
\end{eqnarray}
The density equation gives
\begin{equation}
v_{y,ln}^\mathrm{c} = \frac{-2 n\pi}{l a } v_{z,ln}^\mathrm{s}
\mbox{ and }
v_{y,ln}^\mathrm{s} = \frac{(2 n-1)\pi}{l a } v_{z,ln}^\mathrm{c} .
\end{equation}

The rates of change of the temperature coefficients are obtained from
the non-linear energy equation, Eq.~(\ref{Eq:dTdt1}), again using
trigonometric orthogonality. The left-hand side
of this equation is $\partial T/\partial t$, which is non-zero in
the approach to the steady state. Hence one can update the
temperature field by simple time stepping, with the new temperature
coefficients obtained by adding a constant $\Delta_\mathrm{t} \sim
{\cal O}(10^{-4})$ times the right-hand side to the previous value.

Linear stability analysis reveals that the critical Rayleigh number
is ${\cal R}_\mathrm{c} = 1708$ and the critical wave number is
$a_\mathrm{c} = 3.117$.\cite{Yih77,Drazin81}
For a given Rayleigh number ${\cal R} > {\cal R}_\mathrm{c}  $,
there is a range of wave numbers $a$ that yield steady state solutions.
These are the neutrally stable states.
Of these, some wave numbers are unstable to the cross roll
and other transitions.

%%%%%%%%%%%%%%%%%%%%%%%%%%%%%%%%%%%%%%%%%%%%%%%%%%%%%%%%%%%%%%%%%%%%%%%%
\subsection{Nusselt Number} \label{Sec:Nusselt}

The Nusselt number is the ratio of the total heat flux in convection
to that in conduction at a given Rayleigh number. The heat flux in
conduction is just Fourier's law, $J_\mathrm{E}^\mathrm{cond} = -
\lambda \Delta_{T}$. Since the velocity vanishes at the horizontal
boundaries, the heat flux in convection is purely conductive across
these boundaries. Integrating over a single convection cell, the
Nusselt number is
\begin{eqnarray} \label{Eq:Nu-straight}
{\cal N} & = &
\frac{1}{L_y J_\mathrm{E}^\mathrm{cond}}
\int_{-L_y}^0 \mathrm{d}y \, (-\lambda)
\left. \frac{\partial T^\mathrm{total}(y,z)}{\partial z}
\right|_{z=\pm 1/2}
\nonumber \\
& = & 1 - \sum_{n=1}^N T_{0n}^\mathrm{s} 2 n \pi (-1)^n .
\end{eqnarray}
The temperature in the integrand is the sum of the conductive
temperature field plus the convective perturbation,
Eq.~(\ref{Eq:TB}). The conductive part gives rise to the first term,
1, and the convective terms involving $\sin 2 n \pi z$ and $l=0$
give rise to the remainder.

%%%%%%%%%%%%%%%%%%%%%%%%%%%%%%%%%%%%%%%%%%%%%%%%%%%%%%%%%%%%%%%%%%
\subsection{Algorithms}

\subsubsection{Stable States}  \label{Sec:algo-strt-1}

The Fourier expansion of the temperature and velocity fields
in conjunction with the hydrodynamic equations given above
were used to develop two different algorithms
for ideal straight roll convection.
The first algorithm was used to characterise
neutrally stable straight roll convection.
For each such steady state,
single time quantities
such as the temperature and velocity fields,
the heat flow, Eq.~(\ref{Eq:Nu-straight}),
and the static  part of the total entropy,
Eq.~(\ref{Eq:stot-st}), were determined.
In this first ideal straight roll case,
a  wave number $a$
typically in the neutrally stable range $[2,10]$ was chosen,
with $L \approx N \approx 10$.
Neutrally stable means that a steady state ideal straight roll
solution exists at that wave number and Rayleigh number.
This solution might be unstable to perturbations,
either to straight roll states with a different wave numbers,
or to other convecting patterns.

The algorithm proceeded using the equations give above
with simple time stepping,
$T_{ln}^\mathrm{s/c}(t+\Delta_t) = T_{ln}^\mathrm{s/c}(t)
+ \Delta_t \dot T_{ln}^\mathrm{s/c}(t)$.
Usually, the initial point
was chosen as a small non-zero value in some low order modes,
for example
$ T^\mathrm{s}_{0,1} =  T^\mathrm{c}_{1,1} = 10^{-3}$.
No changes to the results were observed
using other starting points.
In some cases, particularly at higher Rayleigh numbers
or toward the extremities of the neutrally stable range,
a previously converged steady state solution
at a nearby wave number or Rayleigh number
was used as the starting point.
When converged, the final steady state
represented ideal straight roll convection
parallel to the $y$-axis
of wavelength $\Lambda = 2\pi/a$.
Most of the power was in the fundamental mode $a$,
with the next most prominent mode being $3a$.
In this type of calculation the fixed wave number
determines the final steady state.
It is most useful for obtaining thermodynamic properties
as a function of the  steady state wave number,
for example, the heat flux, the static part of the entropy,
and the velocity fields.

The Nusselt number  was monitored
and used to halt  the iterative procedure
when its relative change was less than $10^{-5}$.
For many of the results
reported below, $N=10$ and $L=10$.
Some tests were carried out with
up to $N=16$ and $L=16$. By comparison, Busse\cite{Busse67}
used up to $L+N=12$.
In general, three- or four- figure agreement was obtained between the
present results and those of Busse
for Rayleigh numbers up to 30,000. % 19 Jan 10/1

%%%%%%%%%%%%%%%%%%%%%%%%%%%%%%%%%%%%%%%%%%%%%%%%%%%%%%%%%%%%%%%%%%
\begin{table}
\begin{center}
\caption{  % Papers\Current\Heat Flux\Convection Notes\Busse.xls
Measured (silicone oil, ${\cal P} = 930$) and computed velocity
amplitudes, ($\mu$m/s), for the first three harmonics$^\dag$ at $a=
3.117$ and two Rayleigh numbers. } \label{t1}
\begin{tabular}{c c c c c c }
\hline\noalign{\smallskip}
 & Measured\cite{Dubois78}& Busse\cite{Dubois78} & Present
& Measured\cite{Dubois78} & Present \\
& \multicolumn{3}{c}{${\cal R} = 3416$}
& \multicolumn{2}{c}{${\cal R} = 11,391$} \\
%\cline{2-4}  \cline{5-6}
\noalign{\smallskip}\hline\noalign{\smallskip}
$V_y^1$ & 132$\pm$4 & 133 & 137.6 & 337$\pm$10 & 355.1 \\ \noalign{\smallskip}
$V_y^2$ & 5.3$\pm$0.5$^\ddag$& 5$^\ddag$   & 5.1$^\ddag$
& 13.7$\pm$1 & 13.0 \\ \noalign{\smallskip}
$V_y^3$ & 1.5$\pm$0.3& -   & 1.2   & 19$\pm$1   &  18.3 \\ \noalign{\smallskip}
$V_z^1$ & 145$\pm$5 & 138 & 140.6 & 340$\pm$10 & 363.0 \\ \noalign{\smallskip}
$V_z^2$ & 0         & 0   & 0     & 1.7$\pm$2  &   0 \\ \noalign{\smallskip}
$V_z^3$ & 4$\pm$  0.4& 3.8 & 3.9   &  58$\pm$4  & 60.2 \\ \noalign{\smallskip}
\noalign{\smallskip}\hline
\end{tabular}
\end{center}
%\noindent
\flushleft
$^\dag$The $V_z$ are  at $z^{**}=0$,
and the $V_y$ are  at $z^*=0.28$.\\
$^\ddag$At $z^*=0$.
\end{table}
%%%%%%%%%%%%%%%%%%%%%%%%%%%%%%%%%%%%%%%%%%%%%%%%%%%%%%%%%%%%%%%%%%

The amplitudes of the first three harmonics of the velocity field in
convection have been measured by Dubois and Berg\'e
for a silicone oil (${\cal P}= 930$)
constrained at the critical wavelength.\cite{Dubois78,FN1}
Their results are shown in Table~\ref{t1},
together with their reports of
the results of Busse's calculations, and with the results of the
present calculations, which should be equivalent to those of Busse
(apart from the non-linear influence of the greater number of modes
used here). There is quite good agreement between all three, which
confirms both the validity of the present computational algorithm
and the applicability of the hydro\-dynamic model to the experimental
situation.

%%%%%%%%%%%%%%%%%%%%%%%%%%%%%%%%%%%%%%%%%%%%%%%%%%%%%%%%%%%%%%%%%%%%
\subsubsection{Conduction--Convection Transition} \label{Sec:cond-conv}

In the second type of straight roll calculation,
a small wave number was chosen as the fundamental,
$a \approx$ 0.2--0.5,
and a large number of modes were used, $L \approx$ 60--100
and $N \approx 10$.
A number of different initial states were tested
including uniform distributions
as well as Gaussian distributions of the temperature coefficients,
and white noise.
It was found that the system converged to a straight roll steady
state that was an odd harmonic of the small wave number,
$ \overline a  = (2 \overline l + 1) a$.
(The odd harmonic is demanded by the Boussinesq symmetry.)
The final mode was identified from the power spectrum.

It was confirmed that the properties of the convecting state
given by this second algorithm
were equal to those given by the first algorithm
with fundamental wave number equal to $ \overline a $.

This second type of calculation
modeled the conduction to  straight roll convection transition.
Whereas the first calculation tells the possible
straight roll steady states at a given Rayleigh number,
the second calculation tells the most likely
straight roll steady state
that results  from a transition directly from the conducting state
at a given Rayleigh number.
The modal power,  the sub-system entropy,
and the reservoir entropy
were monitored as a function of time during the transition.

At each Rayleigh number
6--19 independent trials  were performed.
The most likely wave number $\overline a$ for each trial was recorded
(this is the most likely out of all possible wave numbers $l a$),
as well as the average of these over all the trials,
$\left < \overline a \right>$.

It is important to note that the outcome of
this second algorithm
does not refer to the most likely (or average) state,
but rather to the most likely (or average) transition from the conducting state.
This appears to be a general point that
will be confirmed in the results given below:
for non-equilibrium systems,
one cannot speak of \emph{the} most likely phase,
but only of
%the phase most preferred relative to the present phase,
the optimum transition from the current phase.

%%%%%%%%%%%%%%%%%%%%%%%%%%%%%%%%%%%%%%%%%%%%%%%%%%%%%%%%%%%%%%%%%%%%%%%%%%
%
\section{Cross Roll State} \label{Sec:algo-cross}
%\setcounter{equation}{0} \setcounter{subsubsection}{0}
%
%%%%%%%%%%%%%%%%%%%%%%%%%%%%%%%%%%%%%%%%%%%%%%%%%%%%%%%%%%%%%%%%%%%%%%%%%%

%%%%%%%%%%%%%%%%%%%%%%%%%%%%%%%%%%%%%%%%%%%%%%%%%%%%%%%%%%%%%%%%%%
\subsection{Hydrodynamic Equations and Conditions}

This section sets out the computational algorithm that is used
for cross roll convection.
It is supposed that the system is periodic in the
$x$ and $y$ directions,
$T(x,y,z) = T(x+m\Lambda_x,y+n\Lambda_y,z)$,
$m,n = \pm 1, \pm 2, \ldots$
The wavelength is twice the width of an individual roll,
as they come in counter-rotating pairs,
$\Lambda_x = 2L_x$ and  $\Lambda_y = 2L_y$.
The wave numbers are given by
$\Lambda_x = 2\pi/a_x$,
and $\Lambda_y = 2\pi/a_y$.

The four hydrodynamic equations for the four fields,
(temperature and three velocity components)
are explicitly
\begin{equation}
0 =
\frac{\partial v_x}{\partial x}
+ \frac{\partial v_y}{\partial y}
+ \frac{\partial v_z}{\partial z} ,
\end{equation}
\begin{eqnarray} \label{Eq:dTdt}
\frac{\partial T}{\partial t}
& = &
v_z
+ \frac{\partial^2 T }{\partial x^2}
+ \frac{\partial^2 T }{\partial y^2}
+ \frac{\partial^2  T}{\partial z^2}
\nonumber \\ && \mbox{ }
- v_x \frac{\partial T}{\partial x}
- v_y \frac{\partial T}{\partial y}
- v_z \frac{\partial T}{\partial z} ,
%\nonumber \\ &&
\end{eqnarray}
\begin{equation} \label{Eq:Tvx}
0 =
{\cal R} \frac{ \partial  T }{ \partial x}
+
\nabla^{2} \left[ \frac{ \partial v_z }{ \partial x}
- \frac{ \partial  v_x}{ \partial z} \right] ,
\end{equation}
and
\begin{equation} \label{Eq:Tvy}
0 =
{\cal R} \frac{ \partial  T}{ \partial y}
+
\nabla^{2} \left[ \frac{ \partial  v_z }{ \partial y}
- \frac{ \partial v_y }{ \partial z} \right] .
\end{equation}
Taking the $x$ derivative of the penultimate equation,
the $y$ derivative of the final equation,
using the density equation,
and adding them together gives,
\begin{eqnarray} \label{Eq:Tvz}
0 & = &
{\cal R} \left[  \frac{\partial^2 }{ \partial x^2}
+  \frac{\partial^2 }{ \partial y^2} \right] T
+ \left[  \frac{\partial^2 }{ \partial x^2}
+  \frac{\partial^2 }{ \partial y^2}
+  \frac{\partial^2 }{\partial z^2} \right]^2 v_z
\nonumber \\ & \equiv &
{\cal R} \nabla_\parallel^2  T
+ \nabla^{2}  \nabla^{2}  v_z .
\end{eqnarray}
In all these equations,
the temperature and velocity fields are all functions of the position,
${\bf r} = \{x,y,z\}$.
In the steady state they are not functions of time,
but in a transition between states they are.
%The partial time derivative of the temperature
%has been retained on the left-hand side of the energy equation
%as this will be used as the basis for an iterative procedure
%to obtain the steady solution,
%as well as to obtain the evolution of the fields during a transition.

%%%%%%%%%%%%%%%%%%%%%%%%%%%%%%%%%%%%
\subsubsection{Boundary Conditions}

The temperature that appears here is the perturbation due to
convection. That is, the conductive solution has been subtracted
from these equations.
This means that the temperature perturbation must vanish
at the upper and lower boundaries,
\begin{equation}
T(x,y,\pm 1/2) = 0 .
\end{equation}
Since no fluid can cross the boundaries one must also have
\begin{equation}
v_z(x,y,\pm 1/2) = 0 .
\end{equation}
The boundaries are solid walls at which the fluid sticks,
so that one also has
\begin{equation}
v_x(x,y,\pm 1/2) = v_y(x,y,\pm 1/2) = 0 .
\end{equation}
Since this last equation implies that
$\partial v_x(x,y,\pm 1/2)/\partial x =
\partial v_y(x,y,\pm 1/2)/\partial y = 0$,
the density equation implies that
$\left. \partial v_z(x,y,z)/\partial z \right|_{z=\pm 1/2} = 0$.

%%%%%%%%%%%%%%%%%%%%%%%%%%%%%%%%%
\subsubsection{Symmetry}

%As mentioned above,
%the convection  pattern is periodic in the $x$-  and $y$-directions
%with wavelengths $\Lambda_x = 2L_x$ and $\Lambda_y = 2L_y$,
%respectively.
The fundamental convection cell,
$-L_x \le x \le L_x$ and $-L_y \le y \le L_y$,
contains two counter rotating rolls in each direction.
Hence there is mirror plane symmetry at $x=0$ and at $y=0$.
This means that
\begin{eqnarray}
T(x,y,z) & = & T(-x,y,z) = T(x,-y,z) ,
\nonumber \\
v_x(x,y,z) & = & -v_x(-x,y,z) = v_x(x,-y,z) ,
\nonumber \\
v_y(x,y,z) & = & v_y(-x,y,z) = -v_y(x,-y,z) ,
\nonumber \\
v_z(x,y,z) & = & v_z(-x,y,z) = v_z(x,-y,z) .
\end{eqnarray}
These mean that the expansion for the temperature
must be even in the lateral coordinates,
and so must consist of terms like
$T_{qp}(z) \cos(q a_x x) \cos(p a_y y)$, $q$ and $ p $
being non-negative integers.
A similar expansion holds for $v_z$.
Obviously for $v_x$ and $v_y$ the respective cosine is replaced by a sine.

The Boussinesq symmetry refers to the reflection symmetry within a roll.
For ${\bf r} = \{x,y,z\}$,
define ${\bf r}^\dag = \{L_x-x,L_y-y,-z\}$.
One must have
\begin{equation}
T({\bf r}) = -T({\bf r}^\dag)
, \mbox{ and }
{\bf v}({\bf r}) = -{\bf v}({\bf r}^\dag) .
\end{equation}
These basically say that the convective temperature perturbation
at the top of an up draught
must be equal and opposite to that at the bottom of a down draught.
It may be confirmed that the hydrodynamic equations in
Boussinesq approximation given above satisfy this symmetry.

%%%%%%%%%%%%%%%%%%%%%%%%%%%%%%%%%%%%%%%%%%%%%%%%%%%%%%%%%%%%%%%%%%%%%
\subsection{Fourier Expansion}

In view of the Boussinesq symmetry
and the facts that
$\cos(q a_x (L_x-x)) = (-1)^q \cos(q a_x x)$
and  $\cos(p a_y (L_y-y)) = (-1)^p \cos(p a_y y)$,
one sees that $T_{qp}(z)$
must be an odd function of $z$ if $q+p$ is even,
and it must be an even function of $z$ if $q+p$ is odd.
Since $T(x,y,\pm 1/2) = 0$,
and since $\sin(2n\pi z)$ and $\cos((2n+1)\pi z)$
vanish at $z =\pm 1/2$,
the expansion for the temperature is
\begin{eqnarray}
T({\bf r}) & = &
\sum_{q=0}^{Q} \sum_{p=0}^{P}  \sum_{n=0}^{N}
T_{qpn} \cos q a_x x \cos p a_y y
\nonumber \\  && \mbox{ }\times
\left\{ \begin{array}{ll}
\sin 2 n \pi z , & q+p \mbox{ even}, \\
\cos 2 n' \pi z ,  & q+p \mbox{ odd},
\end{array} \right.
\end{eqnarray}
where $n' \equiv (2n+1)/2$,
and $T_{qp0} = 0$ if $q+p$ is even.
%$T_{qmn}^\mathrm{s} = 0$ if $q+m$ is odd or if $n=0$, a
% nd $T_{qmn}^\mathrm{c} = 0$ if $q+m$ is even.
The vertical component of velocity has a similar expansion
\begin{eqnarray}
v_z({\bf r}) & = &
\sum_{q=0}^{Q} \sum_{p=0}^{P}  \sum_{n=0}^{N}
v^z_{qpn} \cos q a_x x \cos p a_y y
\nonumber \\  && \mbox{ }\times
\left\{ \begin{array}{ll}
\sin 2 n \pi z , & q+p \mbox{ even}, \\
\cos 2 n' \pi z ,  & q+p \mbox{ odd} .
\end{array} \right.
\end{eqnarray}

The $x$ component of velocity has the expansion
\begin{eqnarray} \label{Eq:vx-expnsn}
v_x({\bf r}) & = &
\sum_{q=0}^{Q} \sum_{p=0}^{P}  \sum_{n=0}^{N}
v^x_{qpn} \sin q a_x x \cos p a_y y
\nonumber \\  && \mbox{ }\times
\left\{ \begin{array}{ll}
\cos 2 n' \pi z ,  & q+p \mbox{ even} , \\
\sin 2 n \pi z , & q+p \mbox{ odd},
\end{array} \right.
\end{eqnarray}
with $v^x_{0pn} = 0$.
Similarly
\begin{eqnarray} \label{Eq:vy-expnsn}
v_y({\bf r}) & = &
\sum_{q=0}^{Q} \sum_{p=0}^{P}  \sum_{n=0}^{N}
v^y_{qpn} \cos q a_x x \sin p a_y y
\nonumber \\  && \mbox{ }\times
\left\{ \begin{array}{ll}
\cos 2 n' \pi z ,  & q+p \mbox{ even} ,\\
\sin 2 n \pi z , & q+p \mbox{ odd},
\end{array} \right.
\end{eqnarray}
with $v^y_{q0n} = 0$.
These latter two expansions arise from the Boussinesq symmetry
and the vanishing of the velocity at $z = \pm 1/2$.

%%%%%%%%%%%%%%%%%%%%%%%%%%%%%%%%%%%%%%%%%
\subsubsection{$z$-Component of the Velocity}

Inserting the expansions for the temperature and the $z$-component
of the velocity into
Eq.~(\ref{Eq:Tvz}) and setting the coefficients of each term to zero
gives the particular solution
\begin{equation}
v^\mathrm{zp}_{qpn}
=
\frac{{\cal R} [ (q a_x)^2 + (p a_y)^2 ]
}{[ (q a_x)^2 + (p a_y)^2 + {\cal M}_n^2 ]^2}
T_{qpn}
\end{equation}
where
\begin{equation}
{\cal M}_n
\equiv \left\{ \begin{array}{ll}
2 n \pi , & q+p \mbox{ even}, \\
(2 n+1) \pi ,  & q+p \mbox{ odd} .
\end{array} \right.
\end{equation}
Note the distinction between the superscript p, for particular,
and the subscript $p$, an integer index.

The homogeneous solution satisfies
$\nabla^2 \nabla^2 v_z^\mathrm{h}({\bf r}) = 0$.
It has the expansion
\begin{eqnarray}
v_z^\mathrm{h} ({\bf r}) & = &
%\sum_{q=0}^{Q} \sum_{p=0}^{P}  \sum_{n=0}^{N}
%v^\mathrm{zh}_{qpn} \cos q a_x x \cos p a_y y
% \\ \nonumber && \mbox{ }\times
%\left\{ \begin{array}{ll}
%\sin 2 n \pi z , & q+p \mbox{ even}, \\
%\cos 2 n' \pi z ,  & q+p \mbox{ odd} .
%\end{array} \right.
%\nonumber \\ & = &
\sum_{q=0}^{Q} \sum_{p=0}^{P}
\cos q a_x x \cos p a_y y
 \nonumber \\ && \mbox{ }\times
\left\{ \begin{array}{ll}
f_{qp}^\mathrm{zs}(z) , & q+p \mbox{ even}, \\
f_{qp}^\mathrm{zc}(z) ,  & q+p \mbox{ odd} .
\end{array} \right.
\end{eqnarray}
Here it may be readily verified that
the odd homogeneous solution is
\begin{equation}
f_{qp}^\mathrm{zs}(z) =
A_{qp} \sinh(\alpha_{qp} z) + B_{qp} z \cosh(\alpha_{qp} z) ,
\end{equation}
and that the even homogeneous solution is
\begin{equation}
f_{qp}^\mathrm{zc}(z) =
A_{qp} \cosh(\alpha_{qp} z) + B_{qp} z \sinh(\alpha_{qp} z) ,
\end{equation}
with
\begin{equation}
\alpha_{qp} =  \sqrt{ (q a_x)^2 + (pa_y)^2 } .
\end{equation}
Even and odd in this context refer to the parity
with respect to $z$.

With $v_z({\bf r})  = v_z^\mathrm{p}({\bf r})  + v_z^\mathrm{h}({\bf r}) $,
the four boundary conditions,
$v_z(x,y,\pm 1/2) = 0$
and $\left. \partial v_z(x,y,z)/\partial z \right|_{z=\pm 1/2} = 0$,
determine the coefficients $A_{qp}$ and $B_{qp}$
in the even and odd cases.
It is straightforward to use the orthogonality properties
of the trigonometric functions to obtain the Fourier coefficients,
$v^z_{qpn}$.

%%%%%%%%%%%%%%%%%%%%%%%%%%%%%%%%%%%%%%%%%%%%%%%%%%%%%%%%
\subsubsection{$x$- and $y$-Components of the Velocity}

The lateral components of the velocity
can be written in the form
\begin{eqnarray}
v_x({\bf r}) & = &
\sum_{q=0}^{Q} \sum_{p=0}^{P}
f^x_{qp}(z) \sin q a_x x \cos p a_y y ,
\end{eqnarray}
and
\begin{eqnarray}
v_y({\bf r}) & = &
\sum_{q=0}^{Q} \sum_{p=0}^{P}
f^y_{qp}(z) \cos q a_x x  \sin p a_y y .
\end{eqnarray}
%In these $f^x_{0p}(z) = f^y_{q0}(z) = 0$.
%The coefficient functions $f^x_{qp}(z) $ and $f^y_{qp}(z) $
%are even in $z$ if $q+p$ is even,
%and odd in $z$ if $q+p$ is odd.
In view of Eqs~(\ref{Eq:Tvx}) and (\ref{Eq:Tvy}),
one has
\begin{equation}
f^x_{qp}(z) = q a_x  f_{qp}(z)
, \mbox{ and }
f^y_{qp}(z) = p a_y  f_{qp}(z) .
\end{equation}
Inserting these into the density equation,
$\nabla \cdot {\bf v}({\bf r}) = 0$,
and equating the lateral coefficients
term by term yields
\begin{eqnarray}
f_{qp}(z)
& = &
\frac{1}{(q a_x)^2 + (p a_y)^2 }
\sum_{n=1}^N  v_{qpn}^z
\nonumber \\ &&  \times %\mbox{ }
\left\{ \begin{array}{ll}
(-2n \pi ) \cos(2n \pi z) , & q+p \mbox{ even}, \\
(2n'\pi )\sin(2n'\pi z) , & q+p \mbox{ odd}.
\end{array} \right.
\end{eqnarray}
The boundary condition
$ {\bf v}(x,y,\pm 1/2) = {\bf 0}$
is automatically satisfied, having already been invoked
in the solution of the $z$-component of the velocity.
These solutions for the horizontal velocity
can now be projected onto the original $z$-expansions,
Eq.~(\ref{Eq:vx-expnsn}) and Eq.~(\ref{Eq:vy-expnsn}).

%%%%%%%%%%%%%%%%%%%%%%%%%%%%%%%%%%%%%%%%%%%%%%%%%%%%%%%%%%%%%%%%%%
\subsection{Algorithm} \label{Sec:algo-cross-detail}

This cross roll algorithm was used to model the cross roll transition
from a steady straight roll state with wave number $a_y$
to an orthogonal straight roll state with wave number
$\overline a_{x}$.
As in the first form of the ideal straight roll algorithm
described in \S\ref{Sec:algo-strt-1},
$a_y$ was fixed typically in the neutrally stable range $\approx [2,10]$,
with $P = 6 $ and $N=6$.
(The $P$ used here replaces the $L$ used there.)
This gave an adequate
description of the straight rolls in comparison
with the benchmark $L=N=10$ calculations reported there.
A small $x$-wave number, $a_x \approx $ 0.1--0.5,
and a large number of modes, $Q \approx$ 60--150 were used.

For the initial state,
a steady straight roll state of  wave number $a_y$
generated by the first algorithm was used.
In most cases,
the value chosen for the initial wave number
lay toward one of the extremities of the range of neutrally stable wave numbers
for the Rayleigh number
in the expectation of a transition to an intermediate wave number.
The steady state temperature field
was perturbed by adding
an independent random number to each $T_{qpn}$
(white noise).
The amplitude of the noise was proportional
to the square root of the total power in the initial steady state.
Typically, following the addition of the perturbation,
the total power in the $y$-modes increased by about 10\%,
and the total power in the $x$-modes was twice as great
as that in the $y$-modes,
with $\approx 15$ times as many $x$-modes as $y$-modes.
A cross roll transition  usually occurred
to ideal straight rolls in the $x$-direction,
with $\overline a_{x} = (2 \overline q +1 ) a_x \approx $ 3--4.5.
Simple time stepping was used to obtain
the evolution of the temperature field.
The modal power, the sub-system entropy, and the reservoir
entropy were monitored during the transition.

In some cases the computational burden was reduced
by only allowing fundamental modes within
a window about the likely outcome.
That is,
the temperature and velocity coefficients were set to zero
unless they corresponded to an integer multiple (including zero)
of a wave number within the window.
Typically, a window 1--2 wave numbers wide
halved the number of possible Fourier modes,
and reduced the computer time by almost an order of magnitude.
Tests with and without the window
showed that it had little or no effect on the computed transitions.
For example,
at a Rayleigh number of ${\cal R} = 5000$,
and for an initial wave number of $a_y=4.3$,
the average final wave number following the cross roll transition
was $\left< \overline a_x \right> = 3.32 \pm 0.05$
using a wave number step of  $a_x = 0.15$ and no window,
it was $\left< \overline a_x \right> = 3.32 \pm 0.06$
with  $a_x = 0.16$ and a window $[2.5,4.5]$,
and it was $\left< \overline a_x \right> = 3.21 \pm 0.04$
with $a_x = 0.11$ and a window $[2.9,4.0]$.
The standard deviation of the transitions
appeared to depend upon the wave number step,
with it being larger than or equal to the wave number step.

These equations for cross roll
convection and their computational implementation
have been tested by setting $Q=0$ and comparing
the steady state results with those obtained
with the independent straight roll program.

%%%%%%%%%%%%%%%%%%%%%%%%%%%%%%%%%%%%%%%%%%%%%%%%%%%%%%%%%%%%%%%%%%
%
\section{Convection Theory and Experiment} \label{Sec:Results}
%
%%%%%%%%%%%%%%%%%%%%%%%%%%%%%%%%%%%%%%%%%%%%%%%%%%%%%%%%%%%%%%%%%%

%%%%%%%%%%%%%%%%%%%%%%%%%%%%%%%%%%%%%%%%%%%%%%%%%%%%%%%%
%\noindent
\begin{figure}[t!]
\centerline{ \resizebox{8.5cm}{!}{ \includegraphics*{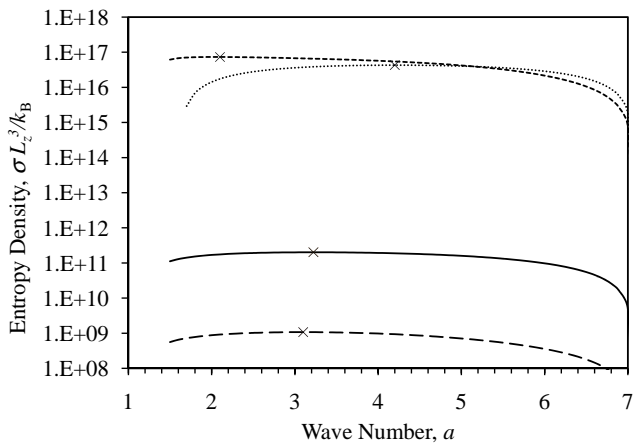}
} }
% \Projects\Convection\2012\Apr12.xlsx
\caption{\label{ScptvsR}
Components of the convective entropy density (log scale)
%as a function of the wave number
at ${\cal R} = 5000$.
The solid curve is the gravitational contribution,
Eq.~(\ref{Eq:s-st-g}),
the long dashed curve is the negative of the kinetic energy contribution,
Eq.~(\ref{Eq:s-st-ke}),
and the short dashed curve is the negative of the internal energy contribution,
Eq.~(\ref{Eq:s-st-int}).
The dotted curve is the sub-system entropy,
Eq.~(\ref{Eq:sigma-st}).
The crosses mark the maximum of each curve.
%Here and in all figures,
%the parameters are for silicone oil,
%footnote \ref{Dubois78} on p.~\pageref{Dubois78},
%except that the kinematic viscosity is $\nu= 0.1\,$cm$^2$/s,
%and $L_z=0.5\,$cm.
}
\end{figure}
%%%%%%%%%%%%%%%%%%%%%%%%%%%%%%%%%%%%%%%%%%%%%%%%%%%%%%%%%

In this section the results of the convection calculations
are presented using the material properties
for a typical silicone oil with $\nu = 0.1$.\cite{Dubois78,FN1}
Figure \ref{ScptvsR}
shows the three contributions to the convective entropy density
at ${\cal R} = 5000$.
This is the difference in entropy between convection and conduction
calculated using the static part of the reservoir entropy,
as described in \S\ref{Sec:s1}.
The figure also shows the sub-system entropy density
(i.e.\ the difference in the sub-system entropy
between the convecting state and the conducting state).
All of the wave numbers
in this and the following figures correspond to hydrodynamic steady states,
with the low wave number end of the curves signifying
the limit to the steady state ideal straight roll
solutions to the Boussinesq equations.
These are the so-called neutrally stable states
for which steady state straight roll solutions exist.
The central region of the neutrally stable region
is absolutely stable,
(i.e.\ perturbations decay and the original
straight roll state remains).
The extreme wave numbers toward the boundaries
of the neutrally stable region are generally unstable to perturbations,
which result in either a new straight roll state with a different wave number,
as might occur via the zig-zag or cross roll transition,
or else another type of convective pattern.

It can be seen that
the entropy due to the convective temperature field itself,
the internal energy contribution Eq.~(\ref{Eq:s-st-int}),
is about six orders of magnitude greater than
the entropy directly due to gravity,
which in turn
is about two orders of magnitude greater
than the entropy due to the kinetic energy.
These results are typical for the whole range
of Rayleigh numbers.
Hence due to this dominance
it makes no difference whether one discusses
the full static form of the convective entropy
or just the  internal energy contribution.
The sub-system entropy itself,
Eq.~(\ref{Eq:sigma-st}),
is comparable in magnitude to the static convective entropy,
but it is positive.
It decreases in magnitude
approaching the limits of the range of steady state solutions
and actually becomes negative
at the low wave number end in most cases.

%%%%%%%%%%%%%%%%%%%%%%%%%%%%%%%%%%%%%%%%%%%%%%%%%%%%%%%%
%\noindent
\begin{figure}[t!]
\centerline{ \resizebox{8.5cm}{!}{ \includegraphics*{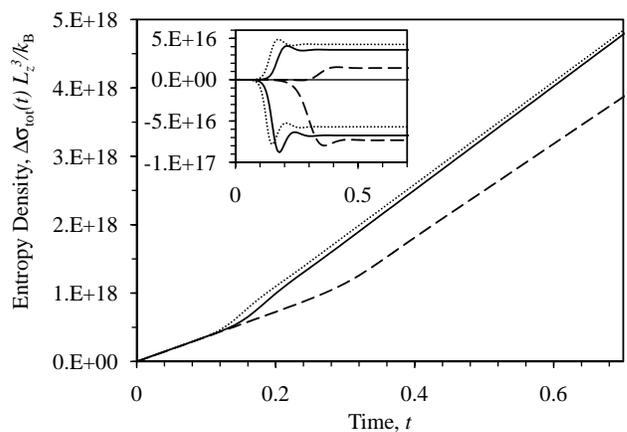}
} }
% \Projects\Convection\2012\Apr12.xlsx
\caption{\label{Sig-st-vst}
Change in entropy density
during a spontaneous straight roll transition from conduction
at ${\cal R} = 5000$.
The solid curves are for $a=3$,
the dashed curves are for $a=2$,
and the dotted curves are for $a=4$.
In the main figure,
the curves are
the change in the total entropy density,
Eq.~(\ref{Eq:Stot-trans}).
In the inset,
the lower three curves are
the internal entropy part
of the static convection entropy, Eq.~(\ref{Eq:s-st-int}),
the upper three curves
are the sub-system entropy,
Eq.~(\ref{Eq:sigma-st}),
and the horizontal line is a guide to the eye.
}
\end{figure}
%%%%%%%%%%%%%%%%%%%%%%%%%%%%%%%%%%%%%%%%%%%%%%%%%%%%%%%%%

The static convective entropy difference,
which, as shown in Fig.~\ref{ScptvsR}, is  dominated by the  internal energy,
is negative.
One should \emph{not} conclude from this
that the convecting state is thermodynamically unfavorable,
since this would contradict the hydrodynamic calculations,
which show that the convecting states are stable
and arise spontaneously from the conducting state
with an initial perturbation.
It can be seen in Fig.~\ref{Sig-st-vst}
that the change in the total entropy density, Eq.~(\ref{Eq:Stot-trans}),
during a straight roll transition is positive.
This contrasts with the static part of the entropy difference,
Eq.~(\ref{Eq:s-st-int}), which is negative throughout the transition.
The difference in the sub-system entropy,
Eq.~(\ref{Eq:sigma-st}), is mainly positive,
but not during the entire transition.
(The initial data for $a=2$ in the inset,
which can only just be resolved on the scale of the figure,
is negative.)
The slope of the change in total entropy density
is the dissipation
(the rate of entropy production of the reservoirs),
which is proportional to the Nusselt number for each case.
It can be seen that it is constant at longer times
as the sub-system achieves its final structure.
The entropy change of the reservoirs completely dominates
the change in entropy of the total system during the transition.
It is always found that the change in the total entropy density
is positive at each stage of the conduction--convection transition.
This means that there is indeed consistency between hydrodynamic stability
and  the Second Law of Thermodynamics.

%%%%%%%%%%%%%%%%%%%%%%%%%%%%%%%%%%%%%%%%%%%%%%%%%%%%%%%%
%\noindent
\begin{figure}[t!]
\centerline{ \resizebox{8.5cm}{!}{ \includegraphics*{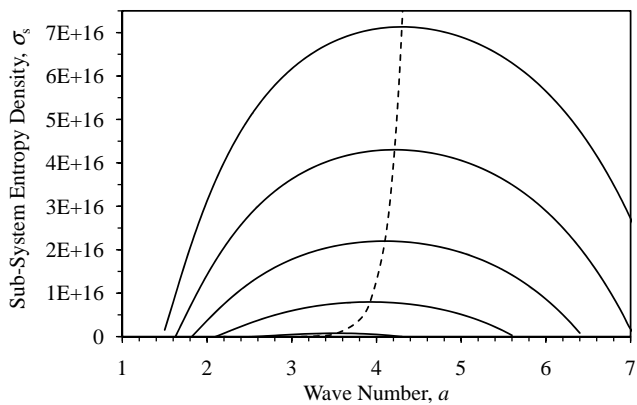}
} }
% \Projects\Convection\2012\Apr12.xlsx
\caption{\label{Ssvsa}
Sub-system entropy density for ideal straight roll convection
for wave numbers in the region of neutral stability,
for Rayleigh numbers from $2000$ (bottom)
to $6000$ (top), in steps of 1000.
The  dashed curve shows the maxima.
}
\end{figure}
%%%%%%%%%%%%%%%%%%%%%%%%%%%%%%%%%%%%%%%%%%%%%%%%%%%%%%%%%

Figure \ref{Ssvsa} shows
the difference in the sub-system entropy
between convection and conduction,
Eq.~(\ref{Eq:sigma-st}),
for ideal straight rolls
as a function of wave number for several Rayleigh numbers.
Steady state straight roll solutions could be obtained
in the range $1708 \le {\cal R} \lesssim 55,000$, %\alts\stackrel{<}{\sim}
with the range of neutral stability
increasing with increasing Rayleigh number
(at least initially).
It can be seen that at a given wave number,
the sub-system entropy density
increases with increasing Rayleigh number,
and that it approaches zero
toward the ends of the stable range.
%It can be negative with small magnitude
%near the lower end of the stable range.
The linear stability analysis of the hydrodynamic equations predicts
that the convective transition occurs at ${\cal R}_\mathrm{c} =1708$
and $a_\mathrm{c} =3.117$.\cite{Yih77,Drazin81}
As the critical  Rayleigh number is
approached from above along the critical wave number,
the sub-system entropy density  approaches zero from above.
Apart from the low wave number end of the steady state range,
the change in sub-system entropy density
from conduction was found to be positive.
As mentioned in connection with Fig.~\ref{Sig-st-vst},
the sub-system entropy density was not always positive
during the approach to the steady state,
and it was also negative in many stable states
toward the low wave number end of the range.

It can be seen in Fig.~\ref{Ssvsa} that the first and the second derivatives
of the sub-system convective entropy density
vanish at the critical wave number and critical Rayleigh number.
This is analogous to behaviour
in equilibrium systems where entropy derivatives vanish
at the critical point.
In convection, it is known that the hydrodynamic fluctuations diverge
at the convective instability
(see Ref.~\onlinecite{Ortiz01} and references therein).
The precise meaning of the vanishing of the sub-system entropy
derivatives at the critical point
and the connection with divergent hydrodynamic fluctuations
is not clear to the present author.

%%%%%%%%%%%%%%%%%%%%%%%%%%%%%%%%%%%%%%%%%%%%%%%%%%%%%%%%
%\noindent
\begin{figure}[t!]
\centerline{ \resizebox{8.5cm}{!}{ \includegraphics*{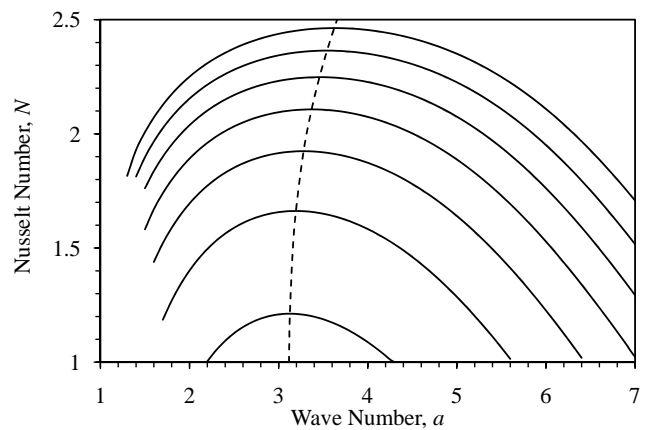}
} }
% \Projects\Convection\2012\Apr12.xlsx
\caption{\label{Nuvsa}
Nusselt number as a function of wave number,
for Rayleigh numbers from $2000$ (bottom)
to $8000$ (top), in steps of 1000.
The maxima are shown by the dashed curve.
}
\end{figure}
%%%%%%%%%%%%%%%%%%%%%%%%%%%%%%%%%%%%%%%%%%%%%%%%%%%%%%%%%

Figure \ref{Nuvsa} shows the Nusselt number
for various Rayleigh numbers
as a function of wave number
over the region of neutral stability.
As in the preceding figures,
the curves represent steady state straight roll convection.
%It can again be seen
%that the neutrally stable range of steady state wave numbers
%increases with increasing Rayleigh number.
There is a well-defined wave number
of maximum heat flux at each Rayleigh number,
and this increases with increasing Rayleigh number.

The significance of this figure relates to the discussion in
\S \ref{Sec:Max-Min-Diss} regarding the Principles of Maximum and
Minimum Dissipation,
which  assert respectively that the optimum non-equilibrium state
is determined by either the greatest or else the least rate
of entropy production.
Since the dissipation is linearly proportional to the Nusselt number,
if one or other of these principles were true,
then the optimum wave number for straight roll convection
at each Rayleigh number
could  be read directly from Fig.~\ref{Nuvsa}.
It can be seen that the wave number of minimum dissipation
occurs at the large wave number boundary of the neutrally stable range,
and that the wave number of maximum dissipation
is near the critical wave number and increases with increasing Rayleigh number.

%%%%%%%%%%%%%%%%%%%%%%%%%%%%%%%%%%%%%%%%%%%%%%%%%%%%%%%%
%\noindent
\begin{figure}[t!]
\centerline{ \resizebox{8.5cm}{!}{ \includegraphics*{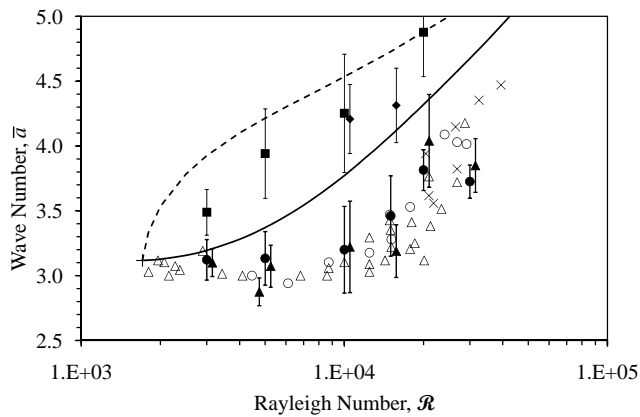}
} }
% \Projects\Convection\2012\Apr12.xlsx
\caption{\label{afvsR}
The ideal straight roll wave number following
a spontaneous transition as a function of Rayleigh number
(log scale).
The open symbols are measured cross roll transitions\cite{Busse71}
and the closed circles and triangles are cross roll  calculations
($P=N=6$, $Q=$100--150, $a_x = $ 0.1--0.2,
error bars give the standard deviation,
with the data shifted horizontally by $\pm$5\%).
The initial constrained state had
a large wave number (triangles),
a medium wave number (crosses),
or a small wave number (circles).
The closed squares ($L=100$ and $a_y = 0.2$)
and closed diamonds ($L=150$ and $a_y = 0.1$)
are the calculated wave number following a conduction--convection transition.
The curves gives the calculated wave number
of maximum Nusselt number (solid  curve)
and  of maximum sub-system entropy density, Eq.~(\ref{Eq:sigma-st}),
(dashed curve).
%The plus symbol %
%terminating the curves
%indicates the critical Rayleigh number and wave number.
}
\end{figure}
%%%%%%%%%%%%%%%%%%%%%%%%%%%%%%%%%%%%%%%%%%%%%%%%%%%%%%%%%

Fig.~\ref{afvsR}
compares several results for the wave number
of ideal straight roll convection:
those that result from spontaneous cross roll transitions,
both experimentally measured\cite{Busse71}
and the present calculations,
those calculated for a direct transition from conduction,
and those calculated to give the maximum Nusselt number
and the maximum sub-system entropy.

The experimentally measured wave numbers
are taken from Fig.~7 of Ref.~\onlinecite{Busse71}.
The experiments were performed by initially constraining
the system in a straight roll convecting state
with a wave number specified by means of a periodic temperature perturbation
(obtained with an intense light source and shadow mask).
Upon removal of the perturbation,
there often occurred a spontaneous transition
via an intermediate cross roll state
to an orthogonal straight roll state
whose wave number is shown in Fig.~\ref{afvsR}.
Only final states that resulted from a cross roll transition
and that are entirely or
predominantly straight rolls are analysed here.
Spontaneous zig-zag transitions to straight roll states,
also occurred but these are not included in the figure.
The majority of the
measurements of Busse and Whitehead\cite{Busse71}
either did not result in
a spontaneous transition, or else did not have a final straight roll
state, or else had too many defects; none of these were analysed.
The measurements of Chen and Whitehead,\cite{Chen68}
did result in ideal straight rolls,
but the data was problematic because they were obtained in a cylindrical cell
and because only the range of final wave numbers
rather than the individual transitions was reported.
Almost all other
measurements in the literature of the convective wavelength could
not be analysed for similar reasons: either no spontaneous
transition occurred, or else the initial state was not a straight roll state,
or else the final state was not a straight roll state,
or else there were many defects in the final state.
\cite{Schmidt38,Deardorff65,Koschmieder66,Rossby69,Krishnamurti70,%
Willis72,Koschmieder74}

It can be mentioned in passing that the measured wave numbers\cite{Busse71}
increase with increasing Rayleigh number,
as can be seen in Fig.~\ref{afvsR}.
Most measurements on convection reported in the literature
show the opposite trend
(see
Refs.~\onlinecite{Koschmieder74b,Normand77,Gollub82b,Berge84,Kolodner86,Cross09}
and also references in Ref.~\onlinecite{Getling98}).
The reason for the difference is that most measurements
are for curved convection rolls,
due either to using a cylindrical cell,
or else to the presence of point defects,
and for these the hydrodynamic equations fix the steady state solutions
to have a wave number that decreases with increasing Rayleigh number.
\cite{Pomeau81,Cross83,Manneville83,Buell86}

Busse and Whitehead\cite{Busse71}
classified the initial wave number as small, medium, or large,
as indicated by the symbols in Fig.~\ref{afvsR}.
The experimental measurements indicate
that the wave number can both increase and decrease
in a spontaneous transition, depending upon
the initial constrained wave number and Rayleigh number.
It is difficult to see a systematic dependence on the initial
wave number in the measured experimental data
(but see Fig.~\ref{axvsay} below).
The measurements (and calculations) show
that there are barriers to changing the straight roll wave number
since continuous evolution of the wave number was not observed,
and once the orthogonal wave number is established
in the intermediate cross roll state, it remains
as the wave number of the final straight roll state.
Consistent with the calculations,
there is a certain width or scatter
in the measured final wave number at each Rayleigh number,
which suggests
that it depends upon the initial wave number, and also
that it depends upon the initial state
or the actual destabilising perturbation.

The experimental data is compared in Fig.~\ref{afvsR}
with the calculated wave numbers
that gives the maximum heat flux
and  the maximum sub-system convective entropy at each Rayleigh number.
There is no real agreement between the observed
final wave numbers and the wave number
that maximises the heat flux or the sub-system entropy.
There is some similarity
in the observed and calculated wave numbers
in that they tend to increase with increasing Rayleigh number.
This similarity is no more than qualitative.
These data provide evidence
that neither the heat flux
nor the sub-system entropy is maximised
in the non-equilibrium state.
Since the static part of the entropy difference is negative,
one can also conclude that it does not determine
the non-equilibrium state.

Results for the final wave number
$\left< \overline a_x \right>$
of the cross roll transition algorithm
described in \S\ref{Sec:algo-cross-detail}
are also shown in Fig.~\ref{afvsR}.
These use $Q=100$ and $a_x = 0.2$,
or $Q=150$ and $a_x = 0.1$.
Both small, $a_y = $ 2.2--2.4,
and large, $a_y =$  4.3--6.0,
wave number initial states were used,
with white noise added as an initial perturbation.
The number of independent trials for each initial wave number
was  5--13.
The bars in the figure signify the standard deviation
of the transition rather than the standard error
of the average final wave number.
In some cases the standard deviation was observed
to depend on the wave number step used in the calculations;
it was generally not less than the  wave number step.
Each calculation was terminated when it was judged
that no further transition would occur,
generally on the basis
that the power in the maximal $x$-mode was clearly dominant.
%and either increasing or else significant and constant,
%and the power in the $y$-modes and in the next two $x$-modes
%was either decreasing or else negligible and constant.
The logarithm of the number of transitions observed from a given
initial wave number to a given final wave number
is directly related to the second entropy of the transition.
It can be seen in Fig.~\ref{afvsR} that there is quantitative
agreement between the  calculated and measured final wave numbers.
This provides additional confirmation
of both the cross roll computer program and algorithm
and also of the applicability of the hydrodynamic model
to the experimental system.

It ought be mentioned that there is the possibility of bias
in the statistics in the present computations
due to the judgement that has to be exercised
in deciding that a transition to an ideal straight roll state
has occurred.
For example,
for the case of a Rayleigh number of ${\cal R} = 5000$
and an initial wave number of $a_y = 6.2$,
16 trials  were run.
Of these, 7 had a clear transition to an ideal straight roll state
by times $t < 4$,
with an average final wave number
$\left< \overline a_x \right> = 3.01 \pm .04$.
The remaining 9 trials exhibited Bloch states,
in which two or more neighboring wave numbers
were dominant with comparable power,
and which were evolving exceedingly slowly.
Five of these Bloch states were terminated at times $t < 4$
and discarded.
The remaining 4 were run until $t=$ 11--13,
with 3 ending up in a clear ideal straight roll state,
with $\left< \overline a_x \right> = 2.77 \pm .07$,
and the remaining 1 being terminated at $t=10.9$
without a clear result
(although it appeared to be converging toward
$\overline a_x = 2.9$).
If one had retained only trials with a clear transition by $t=4$,
then one would have obtained
$\left< \overline a_x \right> = 3.01 \pm .04$.
If one simply averaged over all transitions that had actually been completed,
then the average is
$\left< \overline a_x \right> = 2.94 \pm .05$.
But this last result is biased because slowly evolving cases
are the ones discarded and these would have had a smaller wave number
if they had been allowed to go to completion.
If one removed this bias by weighting the short and long time averages
in the ratio 7:9
(i.e.\ according to the number of Bloch states),
then the average is $\left< \overline a_x \right> = 2.88 \pm .04$.
In summary, in this case (${\cal R} = 5000$, $a_y = 6.2$)
rapid transitions had a larger final wave number
than transitions that took more time.
Only a small minority of all the cases run were terminated without a result,
and so it is likely that this potential bias
does not have a significant effect on the computational results
in Fig.~\ref{afvsR}.
Whether or not it is a potential experimental problem is unclear.

In addition to the cross roll transition,
Fig.~\ref{afvsR} also shows the average
final wave number for the conduction--convection transition,
$\left< \overline a_y \right>$,
whose calculation was described in \S\ref{Sec:cond-conv}.
White noise was used as a perturbation to initiate the transition.
The average final wave number appeared insensitive to the wave number step
($a_y =$ 0.1 or 0.2),
although the standard deviation was smaller for the smaller step.
It was found that in most cases
the system converged to a straight roll steady state
that was an odd harmonic of the small wave number,
$ \overline a  = (2 \overline l + 1) a$,
as demanded by the Boussinesq symmetry.
Beyond ${\cal R} \agt 20,000$,
ideal straight rolls did not result,
or at least there was not a single clearly dominant wave number.
(Some such Bloch states are included in the averages
for ${\cal R} = 15,000$ in Fig.~\ref{afvsR}.)
It can be seen that the wave number resulting from the direct
conduction--convection transition
is larger than that resulting from a cross roll transition.
%The wave number following the conduction--convection transition
%appears to lie between the wave number of maximum Nusselt number
%and the wave number of maximum sub-system entropy.

These results show that the two types of phase transition,
conduction--convection and cross roll,
differ significantly even though both result
in the same type of final state, namely ideal straight rolls.
At a given Rayleigh number,
the ideal straight roll wave number
depends upon whether it results directly from conduction,
or whether it results from a cross roll transition
(and, in the latter case, it depends on the initial
wave number, as shown in Fig.~\ref{axvsay} below).
One can conclude from this
that a single time variational quantity whose optimisation determines
the favored non-equilibrium pattern
either does not exist,
or else has negligible influence compared to
the barriers between the multiple possible patterns.
This is likely to be a general feature of non-equilibrium systems,
and instead of seeking `the' optimum pattern or phase,
one should focus on the optimum transition from a given phase.

%%%%%%%%%%%%%%%%%%%%%%%%%%%%%%%%%%%%%%%%%%%%%%%%%%%%%%%%
%\noindent
\begin{figure}[t!]
\centerline{ \resizebox{8.5cm}{!}{ \includegraphics*{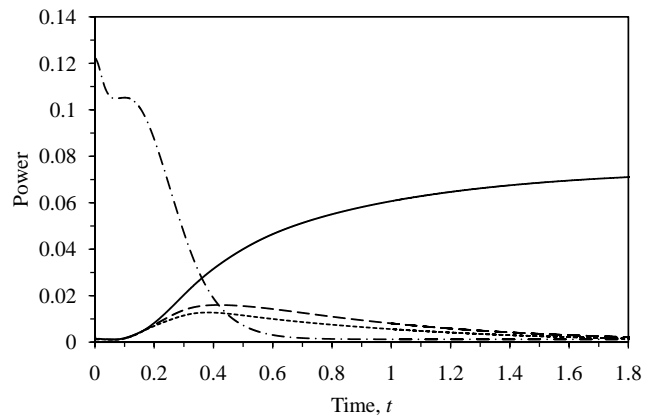}
} }
% \Projects\Convection\2012\Apr12.xlsx
\caption{\label{Pvst}
The calculated modal power
during a cross roll transition
(${\cal R} = 5000$,
$a_y = 1.7$,
$\overline a_x = 3.41$,
$a_x = 0.31$, $Q=80$, and $P=N=6$).
The solid, dashed, and dotted curves are
the three $x$-modes with highest power,
and the dash-dotted curve is the total power
in the $y$-modes.
} \rule{0cm}{.3cm} % strut:
\end{figure}
%%%%%%%%%%%%%%%%%%%%%%%%%%%%%%%%%%%%%%%%%%%%%%%%%%%%%%%%%

The transition between two orthogonal straight roll convective states
via the cross roll intermediate state,
as described in \S\ref{Sec:algo-cross},
can be monitored by the  evolution of the power
in the various modes,
as shown in Fig.~\ref{Pvst}.
In this case the system was initially in a convecting
straight roll steady state
with $a_y = 1.7$,
near the lower end of stable states,
and at $t=0$ white noise was added to all the modes.
This increased the total power in the $y$-modes by about 10\%,
and created almost twice as much power in the $x$-modes as in  the $y$-modes,
spread over thirteen times as many modes.
The power in the $x$-mode $q$ was defined as
$P_{xq} \equiv \sum_n T_{q0n}^2$,
and the total power  in the $x$-modes
was defined as $P_{x} \equiv \sum_q P_{xq} $,
and similarly for the $y$-modes.

The initial white noise perturbation
is meant to model the experimental situation,
although the origin of the noise is unclear.
It is possible that noise originates
from mechanical vibrations
or from temperature inhomogeneities and water flow in the heat baths,
in which cases white noise would likely be appropriate.
%It is a moot point whether using white noise
%as the perturbation of the initial temperature field
%that initiates the cross roll transition
%is the most appropriate approach.
Although the perturbation is small on the scale
of the convective temperature,
it is large on molecular scales,
and so treating it as a thermodynamic fluctuation
(i.e.\ weighting it by the exponential of the entropy)
would be problematic.

%No systematic investigations have been carried
%on the nature of the calculated transitions
%using a different or smaller perturbation.

%%%%%%%%%%%%%%%%%%%%%%%%%%%%%%%%%%%%%%%%%%%%%%%%%%%%%%%%
%\noindent
\begin{figure}[t!]
\centerline{ \resizebox{8.5cm}{!}{ \includegraphics*{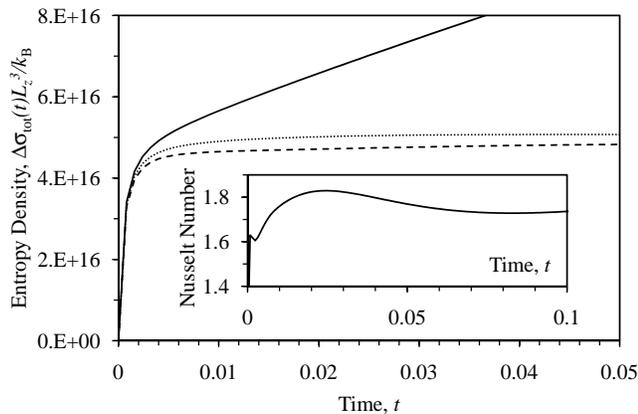}
} }
% \Projects\Convection\2012\Apr12.xlsx
\caption{\label{Sigvst}
Change in entropy density %at short times
during the cross roll transition of the preceding figure.
The solid curve is the change in total entropy density,
Eq.~(\ref{Eq:Stot-trans}),
%which continues to  increase linearly at longer times.
the dashed curve is the change in the internal entropy part
of the static convection entropy, Eq.~(\ref{Eq:s-st-int}),
and the dotted curve is the change in the sub-system entropy,
Eq.~(\ref{Eq:sigma-st}).
The inset shows the Nusselt number.
} \rule{0cm}{.3cm} % strut:
\end{figure}
%%%%%%%%%%%%%%%%%%%%%%%%%%%%%%%%%%%%%%%%%%%%%%%%%%%%%%%%%

Figure~\ref{Pvst} shows that the power in the $x$-modes
grows over time at the expense of the initially stable $y$-modes.
By about $t\approx 0.6$ the $y$-rolls have disappeared,
and by about $t \approx 1.5$ a steady straight roll convecting state
has been established with $\overline a_x = 3.41 = 11 \times 0.31$.
The Nusselt number at $t=1.88$ was ${\cal N} = 2.110$,
which compares well with ${\cal N} = 2.107$
calculated using the ideal straight roll algorithm of \S\ref{Sec:algo-strt}
for ${\cal R} = 5000$ and $a=3.4$.
It can be concluded from the figure that a cross roll transition
from one straight roll state to another has occurred.
It may be called spontaneous in the sense that no constraint was
imposed on the final state (other than that it be an odd integer multiple
of the wave number step) or indeed on whether any transition would occur at all.

Figure \ref{Sigvst}
shows the evolution of the total entropy
during a cross roll transition.
It can be seen that the total entropy monotonically increases in time.
It can also be seen
that the change in the static part of the convective entropy
and the change in the sub-system entropy
are positive during the transition.
The dissipation, which is the rate of change of the reservoir entropy,
is proportional to the Nusselt number.
Although the Nusselt number
is higher for the final state than for the initial state,
it can be seen from the inset that it does not increase monotonically
during the transition.
Since the proposition that the dissipation is maximised
in the optimum non-equilibrium state
implies that the dissipation increases during all stages
of a spontaneous transition,
the data in the inset provides further evidence against the
maximum dissipation idea.

%%%%%%%%%%%%%%%%%%%%%%%%%%%%%%%%%%%%%%%%%%%%%%%%%%%%%%%%
%\noindent
\begin{figure}[t!]
\centerline{ \resizebox{8.5cm}{!}{ \includegraphics*{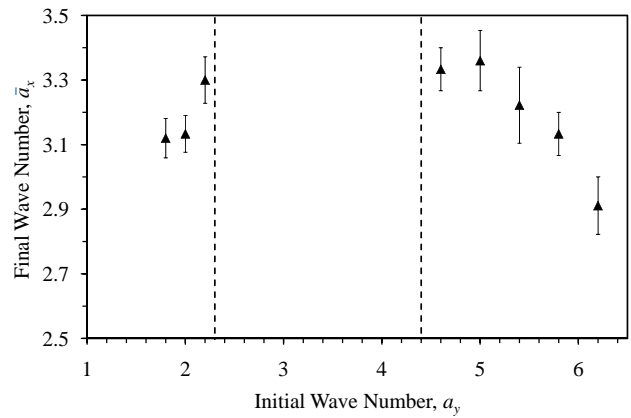}
} }
% \Projects\Convection\2012\Apr12.xlsx
\caption{\label{axvsay}
The calculated final wave number $\overline a_x  $
as a function of the initial wave number  $a_y $
for cross roll transitions at ${\cal R} = 10,000$
($Q=N=6$, $P=100$, and $a_x = 0.2$).
The symbols signify the most likely
final wave number averaged over 6--12 trials,
and the vertical bars show
the standard error on the mean.
The vertical dashed lines bound the region
of stable wave numbers for which no transition occurred.
}
\end{figure}
%%%%%%%%%%%%%%%%%%%%%%%%%%%%%%%%%%%%%%%%%%%%%%%%%%%%%%%%%

Figure \ref{axvsay}
shows the final wave numbers
calculated from a series of cross roll transitions
for different initial wave numbers
at a Rayleigh number of 10,000.
Since only cross roll transitions
were permitted in the computations,
it is possible that the calculated cross roll domain
is larger than the measured one
because unlike the calculations,
in the experiments other transitions can supersede
the cross roll transitions.
The most likely final wave number
for any one trial
is the wave number that,
out of all possible wave numbers  $l a_x$,
has the most power after a transition
has occurred, $\overline a_x = ( 2 \overline l + 1) a_x$.
This varies between trials,
and it is the average of these that is plotted
for each initial wave number.
It can be seen that the final wave number
depends upon the initial wave number.
One aspect of this dependence
is the existence of a region of
stable initial wave numbers for which no transition was observed
(6 wave numbers spanning this region were tested,
with 4--12 trials in each case).
A second aspect is the systematic increase in final wave number
with decreasing initial wave number
for large initial wave numbers.
One can possibly make out the opposite trend
for small initial wave numbers.
This variation of the final wave number
is statistically significant.
It is possible that at least part of the scatter
of the experimental data\cite{Busse71} in Fig.~\ref{afvsR}
can be attributed to a similar
dependence on the initial wave number.
Also, the relatively small change in final wave number,
as well as the opposite trend on either side of the stable region,
are consistent with the experimental data\cite{Busse71} in Fig.~\ref{afvsR}
that appear to show no difference in the final wave number
for systems that were in a small or in a large
wave number initial state.

Another example of this dependence of the final state
on the initial state is the zig-zag transition.
Although not explored here,
it is observed experimentally\cite{Busse71}
that ideal straight rolls initially close to the low wave number
neutrally stable boundary, $a_\mathrm{i}$,
can undergo a spontaneous zig-zag transition
to another ideal straight roll state,
oriented at 45$^\circ$ to the initial rolls,
with final wave number $a_\mathrm{f} = a_\mathrm{i} \surd 2$.
Obviously then this is a clear example
when, at a given Rayleigh number,
the final state depends upon the initial state.
Again the final state
is determined by the thermodynamics of the transition itself
rather than by any single time thermodynamic variational principle
of the final state of ideal straight rolls.

Figure \ref{axvsay} brings the focus
to the transitions rather than the states.
For a given Rayleigh number and initial wave number,
distinct final states
(i.e.\ distinct transitions)
occur with non-negligible probability (not shown).
For example,
in the case of ${\cal R} = 10,000$
and $a_y = 5.4$, 4 distinct transitions
actually occurred in 9 trials,
(${\overline a}_x = 3.4$ occurred 4 times,
${\overline a}_x = 3.0$ occurred 3 times,
${\overline a}_x = 2.6$ occurred once, and
${\overline a}_x = 3.8$ occurred once).
The existence of distinct final states
from which no further transitions occur
signifies that patterns with a given wave number
are locally stable and that there are barriers to further transitions,
which is consistent with the experimental observations.

The data in Fig.~\ref{axvsay}
imply that at a given Rayleigh number,
it is less meaningful to speak of the average
wave number for ideal straight roll convection
%(although such is mathematically well-defined)
than it is to speak of the average
wave number following a transition from a given steady state.
This particular point was already made in
connection with Fig.~\ref{afvsR},
where the distinction between the wave number
resulting from a conduction--convection transition
and from a cross roll transition was discussed.
Even this is a severe simplification
of the full convective transition phenomenon,
since the final wave number of the cross roll transition
depends upon the initial wave number.

%\newpage $\;$ \newpage
%%%%%%%%%%%%%%%%%%%%%%%%%%%%%%%%%%%%%%%%%%%%%%%%%%%%%%%%%%%%%%%%%%
%
\section{Conclusion} \label{Sec:Concl}
%
%%%%%%%%%%%%%%%%%%%%%%%%%%%%%%%%%%%%%%%%%%%%%%%%%%%%%%%%%%%%%%%%%%

%%%%%%%%%%%%%%%%%%%%%%%%%%%%%%%%%%%%%%%%%%%%%%%%%%%%%%%%%%%%
\subsubsection{The Second Law of  Thermodynamics}

For an equilibrium system,
the Second Law of Thermodynamics
provides the variational principle
that determines the optimum state.
The primary statement of the law is that the total entropy increases
during a spontaneous transition.
The corollary of this is that the
total entropy is maximised in the optimum state.
(This is the same as minimising the free energy,
which is minus the temperature times the total entropy.)\cite{TDSM}
%For the discussion that follows,
%it is important to distinguish these two aspects of the Second Law.

The Second Law of Thermodynamics
rests on two distinct bases.
First is quantitative evidence,
originally experimental measurement,
and, in more recent times, computational data.
Second, is reason,
namely Boltzmann's identification
of entropy with the logarithm of the molecular configurations
that comprise a state.
By linking entropy to probability,
Boltzmann provided both an intuitive physical explanation
of why entropy increases during spontaneous transitions,
and also a mathematical prescription for calculating entropy
and its evolution.

Just as important as what the Second Law says
is what it does not say.
Nothing quantitative is said about time.
Implicitly the law gives the direction of time
%as the direction of entropy increase,
but not its speed.
Specifically,
the law says nothing about the rate of entropy increase,
the dissipation.
%This is not a law about non-equilibrium systems
%in the sense that it tells nothing about the rate of entropy increase,
%nor the velocity of flow, nor the speed of motion,
%nor the timing or duration of an event.

There is a class of static systems,
namely those that are hysteretic or metastable,
that illuminate a specific aspect
of the  Second Law of  Thermodynamics
that is relevant to the discussion of heat flow that follows.
Such systems include glasses and slowly relaxing  substances,
macromolecules,
(e.g.\ the shape and conformation of proteins),
and systems stuck beyond the usual phase transition point
(i.e.\ absent nucleation sites).
These systems can be quite stable
(i.e.\ they do not change macroscopically
over experimental time scales)
and in this sense they can quite reasonably be described
as equilibrium systems.
And yet, the current state of such systems can depend
upon their past history.
As far as the Second Law of Thermodynamics is concerned,
it is quite true that the total entropy has increased
during the evolution of such systems to their current state.
But the entropy is not necessarily a maximum in the current state.
One interpretation of such systems
is that they are trapped in a local entropy maximum
(free energy minimum),
and that there are barriers that prevent a transition
to the state of global entropy maximum.
%(One need not quibble here whether a glass is a trapped stationary state,
%or whether it is an extremely slowly evolving state.)

There are two relevant lessons that can be drawn from this example.
The first is the importance of both evidence and reason
for the Second Law of Thermodynamics.
If the law were  based only upon evidence,
then the existence of such hysteretic or metastable systems
would appear to invalidate it,
or at least the corollary that the optimum state
is the state of maximum entropy.
However,
because one has a reasonable physical interpretation of the Second Law,
one can instead identify these cases as exceptions to the rule,
and interpret them in terms of barriers
that preclude transition to the globally optimum state.
The second lesson from this example
is that entropy
as a variational principle can have practical limitations
in determining the actual state of the system.
%However the more fundamental principle that entropy increases
%during spontaneous transitions holds even for
%hysteretic or metastable systems.

%%%%%%%%%%%%%%%%%%%%%%%%%%%%%%%%%%%%%%%%%%%%%%%%%%%%%%%%%%%%
\subsubsection{Specific Conclusions for Heat Flow}

Turning now to the specific results of this paper,
one can draw some definite conclusions for convective heat flow.
%and one can argue from these to some universal principles
%for the behavior of non-equilibrium  and pattern forming systems
%in general.
The Principle of Maximum or Minimum Dissipation
was discussed in the introduction of the paper.
This is the variational principle
that some have postulated for non-equilibrium systems,
namely that
the optimum state is given by either a maximum or a minimum
in the rate of entropy production.
For the case of heat flow, the dissipation
is the heat flux divided by the temperature gradient,
which is essentially the Nusselt number.
The results in Fig.~\ref{afvsR}
for a spontaneous cross roll transition
show that neither the measured\cite{Busse71}
nor the calculated final wave number
correspond to the maximum Nusselt number.
(The wave number of minimum Nusselt number would lie off the scale.)
Similarly the calculated spontaneous  conduction-convection transition
yields a final state that does not correspond
to an extremity of the Nusselt number.
The present results,
as well as earlier results,\cite{Busse67,Busse71}
show that the observed convective patterns
following a spontaneous transition
do not correspond to either the maximum nor to the minimum heat flux.
Hence the superficial conclusion from these quantitative observations
is that neither the maximum dissipation nor the minimum dissipation
yield the optimum non-equilibrium state.

It might be objected that one cannot draw this conclusion solely
from the data because,
as in the case of metastable or hysteretic equilibrium states,
there are barriers that prevent transitions:
a globally optimum state might still exist
%and be given by one or other of the Principles of Extreme Dissipation,
but it is simply not achieved in the experiments or calculations.
In other words,
the violation of the Principles of Extreme Dissipation
might only be a matter of practice rather than of principle.
Continuing the objection,
even if barriers reduce the utility of the Principle in this system,
one or other of the two Principles might still be valid,
and it might be useful for other non-equilibrium systems.

There are two counter arguments to this objection.
The first is quantitative and is illustrated in the inset of
Fig.~\ref{Sigvst}.
There it is shown that the Nusselt number
(equivalently, the heat flux or the dissipation)
evolves non-monotonically during a spontaneous cross-roll transition.
This rules out both Principles of Extreme Dissipation:
the Nusselt number neither always increases nor always decreases
during a spontaneous transition between convective states.
Hence even if there were no barriers between different convective states,
the optimum non-equilibrium state cannot correspond
to an extremum of the Nusselt number.

%Hence in general the dissipation cannot be
%a monotonic function during transitions between non-equilibrium states.
%Hence in general the final non-equilibrium state cannot correspond
%to an extremum of the dissipation
%even when there are no barriers between states.

The second counter argument
is that neither of the Principles of Extreme Dissipation
is supported by reason.
Unlike the Second Law of Thermodynamics,
there is no molecular interpretation
that provides a persuasive picture for
maximal or minimal rates of entropy production.
Hence there is no credible basis
to argue that the results of these experiments and computations
for convection are an exception to a general rule.

Besides ruling out
any Principle of Extreme Dissipation
for convective systems,
the present calculations
allow one to draw another quantitative conclusion for heat convection,
namely that the final state of a spontaneous transition
depends upon the initial state.
This was shown in Fig.~\ref{afvsR},
where different straight roll states resulted,
depending upon whether the initial state was the conducting state,
or whether it was an orthogonal cross-roll state,
and in Fig.~\ref{axvsay}, where it was shown that the final wave number
for the cross roll transition varied with the initial wave number.
A similar dependence is seen experimentally
in the zig-zag transition.\cite{Busse71}
These results imply that for heat convection
a single time thermodynamic quantity
(such as the heat flow, or the sub-system entropy, or the total system entropy)
is practically insufficient to determine the optimum state
following  a spontaneous transition.
The word `practically' here recognises that
it might be the barriers between convective states
that give rise to the dependence on the initial state
and so the insufficiency of a single time quantity
might be one of practice rather than one of principle.
(Just as the insufficiency of entropy for describing a
metastable or hysteretic state
is a limitation in practice that does not rule out
entropy as an equilibrium principle.)

In summary, the present quantitative calculations
for heat convection lead to two specific conclusions.
First,
the Nusselt number
(equivalently, the heat flux or the dissipation)
does not change monotonically
during spontaneous transitions between convecting steady states.
From this it follows that
the Nusselt number cannot provide a variational  principle for heat convection.
And second,
due to the presence of barriers
signified by a dependence of the final state on the initial state,
in practical terms
there cannot be \emph{any} single time
variational principle for heat convection.

%%%%%%%%%%%%%%%%%%%%%%%%%%%%%%%%%%%%%%%%%%%%%%%%%%%%%%%%%%%%
\subsubsection{General Conclusions for Non-Equilibrium Systems}

%What general inferences
%can be drawn from these specific conclusions for heat convection?
The behavior of the Nusselt number
and its equivalence to the dissipation
rules out any general principle for non-equilibrium systems
that asserts
that spontaneous transitions between states
correspond to a monotonic change in the dissipation.
The corollary of this is that
there can be no general Principle of Extreme Dissipation
for determining the optimum non-equilibrium state.
(Non-equilibrium state means both flux and structure.)
The mathematical basis for this general conclusion
is just an axiom of Aristotelian logic:
the single counter example of heat flow in convecting systems
disproves the general theorem.
The physical reason why there can be no principle of extreme dissipation
is outlined below.

Beyond the dissipation,
general conclusion can be drawn about transitions.
The specific results for heat convection obtained here
shifts the focus from non-equilibrium states
to the transitions between them.
This was the conclusion drawn from Figs~\ref{afvsR} and \ref{axvsay},
namely that the observed  non-equilibrium state
is conditional upon the initial state prior to the transition.
This highlights the importance of the second entropy
as the variational principle for non-equilibrium systems,
since it is the two-time thermodynamic quantity
that gives the probability of transitions.

There are two reasons to believe
that in general non-equilibrium systems
are better characterised by their transitions than by their states.
The first argument,
which is no more than suggestive,
is based on the observation that in general
a non-equilibrium state often means a steady pattern
that is defined by the contrast between distinct regions
of fluxes and structure.
%and the relatively sharp edges that delineate them.
It seems reasonable to conclude that each region must represent
a locally stable state,
and there must be barriers inhibiting transformation
from one pattern to another.
For the same reasons as in heat convection,
in practice such barriers negate the efficacy of
single time thermodynamic quantities
compared with the transition entropy.

The second, more rigorous, reason is that
the non-equilibrium systems of interest
can generally be modeled as a thermodynamic gradient imposed
on a sub-system by reservoirs,
and it is the structure and flux of the sub-system
that represents the non-equilibrium state being characterised.
However, the thermodynamic gradient
can also be considered as arising from a spontaneous fluctuation
of the total system,
with the consequent flux being the regression of that fluctuation.
This means that the current state of the sub-system
is really part of the end state of a transition
from the initial fluctuation of the total system
that occurred some time in the past.
In ranking two alternative non-equilibrium states of the sub-system
given the applied thermodynamic gradient,
it is not valid to compare their structural entropies
because one is not comparing two spontaneous
fluctuations of the sub-system.
Likewise, it is not valid to rank them
according to their total entropies,
again because the total entropy \emph{now}
is due not to a fluctuation \emph{now},
but rather to a prior fluctuation of the total system.
(If total entropy was the correct ordinal function,
then the steady state with the greatest dissipation would always
be the preferred state because it would
have the greatest total entropy after sufficiently long times.
Hence the sentence immediately preceding these parenthetical remarks
provides the physical reason for ruling out the Principle of Maximum Dissipation
in general.)
Instead of these one-time thermodynamic functions
(the sub-system entropy, the total entropy, or the dissipation),
one has to compare the two states conditioned on the fact
that the fluctuation of the total system
has already been regressing for some time.
This is why two-time rather than one-time thermodynamic quantities
provide the relevant variational principle
for characterising the non-equilibrium state of the sub-system.

The regression of the total system
obeys the  Second Law of Thermodynamics,
namely that the total entropy increases over time.
Hence the total dissipation is positive for any non-equilibrium state.
(For a steady state of the sub-system,
the total dissipation equals the dissipation of the reservoirs.)
In analysing the possibility of a transition from one steady state
to another,
the Second Law of Thermodynamics demands only that
the total dissipation must continue to be positive.
It does not say that the dissipation has to increase.
%This is the reason that there can be no general Principle
%of Extreme Dissipation.
Even if the entropy of the sub-system for the final state
is lower than that for the initial state,
the transition is permitted provided that
the decrease in sub-system entropy is smaller in magnitude
than the integral of the reservoir dissipation over the time interval
of the transition.
(In such cases, this sets an upper bound on the speed of the transition.)

The second entropy is essentially the logarithm
of the number of molecular configurations
that give the  transition.
Its exponential gives the probability
of a transition from one non-equilibrium state to another.
Hence the second entropy is the two-time thermodynamic quantity
that characterises non-equilibrium states and patterns.

In the present paper the second entropy appeared in two, indirect, ways.
First, the hydrodynamic equations
are equivalent to maximising the second entropy
with respect to the fluxes,
whilst  accounting for the material conservation laws
and equilibrium equation of state
(see Refs~\onlinecite{Landau57,Landau59,Ortiz06,AttardX}
and Ch.~5 of Ref.~\onlinecite{NETDSM}).
Second, the accumulation of trial transitions
between steady convecting states
is equivalent to mapping out the second entropy
conditioned on the initial state.
These two applications of the second entropy
demonstrate its utility.
The specific results obtained here for heat convection
support the general conclusion
that the two-time second entropy
(or its generalisations for non-Markov systems)\cite{NETDSM}
is the variational principle for non-equilibrium systems.

%\newpage %$\;$ \newpage \ \newpage
%{\bf \large References}
%%%%%%%%%%%%%%%%%%%%%%%%%%%%%%%%%%%%%%%%%%%%%%%%%%%%%%%%%%%%%%%%%%%%%%%%%%

%%%%%%%%%%%%%%%%%%%%%%%%%%%%%%%%%%%%%%%%%%%%%%%%%%%%%%%%%%%%%%%%%%%%%%%%%%

\begin{thebibliography}{99}
\bibitem{FN1}
The physical parameters of the silicone oil are:
temperature $T_{00} = 298\,$K,
density $\rho_{00} = 0.960\,$g$\,$cm$^{-3}$,
kinematic viscosity $\nu = 1.056\,$cm$^2\,$s$^{-1}$,
%        nu = nu/10     ! Busse and Whitehead
thermal expansivity $\alpha = 0.96\times 10^{-3}\,$K$^{-1}$,
thermal conductivity
$\lambda = 3.7\times 10^{-4}\,$cal$\,$cm$^{-1}\,$K$^{-1}\,$s$^{-1}$,
specific heat   $c_p^\dag = 0.337\,$cal$\,$g$^{-1}\,$K$^{-1}$,
and thermal diffusivity $\kappa = 1.14\times 10^{-3}\,$cm$^2\,$s$^{-1}$.
From line 1 of Table I  of Ref.~\onlinecite{Dubois78}.
\end{thebibliography}
\end{document}